\documentclass[12pt, prb,  aps, unsortedaddress,superscriptaddress,floatfix,amsmath,amssymb]{revtex4-1}
\usepackage{times}
\usepackage[margin=.85in]{geometry}

\usepackage{graphicx}
\usepackage{amssymb,amsfonts,amsmath}

\usepackage{color}
\usepackage[normalem]{ulem}					% add \sout strickout

\renewcommand\k{\kappa}%accent
\newcommand\la{\langle}
\newcommand\ra{\rangle}
\renewcommand\t{\tau}%accent char
\newcommand\s{\sigma}
\newcommand\ri{\right}
\renewcommand\le{\left}
\renewcommand\th{\theta}

\newcommand\m{\mu}
\renewcommand\t{\tau}%accent char
\renewcommand\th{\theta}%latin char
\newcommand\D{\Delta}

\newcommand\mb{\mathbb}

\begin{document}

\title{Scaling laws governing stochastic growth and division of single bacterial cells}
\author{Srividya Iyer-Biswas}
\affiliation{James Franck Institute and Institute for Biophysical Dynamics, University of Chicago, Chicago, IL 60637}
\author{Charles S. Wright}
\affiliation{James Franck Institute and Institute for Biophysical Dynamics, University of Chicago, Chicago, IL 60637}l
\author{Jonathan T. Henry}
\affiliation{Department of Biochemistry and Molecular Biology, University of Chicago, Chicago, IL 60637}
\author{Klevin Lo}
\affiliation{Department of Biochemistry and Molecular Biology, University of Chicago, Chicago, IL 60637}
\author{Stanislav Burov}
\affiliation{James Franck Institute and Institute for Biophysical Dynamics, University of Chicago, Chicago, IL 60637}
\author{Yihan Lin}
\affiliation{Department of Biology, California Institute of Technology, Pasadena, CA 91125}
\author{Gavin E. Crooks}
\affiliation{Physical Biosciences Division, Lawrence Berkeley National Laboratory, Berkeley, CA 94720}
\author{Sean Crosson}
\affiliation{Department of Biochemistry and Molecular Biology, University of Chicago, Chicago, IL 60637}
\author{Aaron R. Dinner}
\email{dinner@uchicago.edu}
\affiliation{James Franck Institute and Institute for Biophysical Dynamics, University of Chicago, Chicago, IL 60637}
\author{Norbert F. Scherer}
\email{nfschere@uchicago.edu}
\affiliation{James Franck Institute and Institute for Biophysical Dynamics, University of Chicago, Chicago, IL 60637}

%\begin{article}

%\pagebreak
\begin{abstract}
Uncovering the quantitative laws that govern the growth and division of single cells remains a major challenge. Using a unique combination of technologies that yields unprecedented statistical precision, we find that the sizes of individual {\em Caulobacter crescentus} cells increase exponentially in time.  We also establish that they divide upon reaching a critical multiple ($\approx$1.8) of their initial sizes, rather than an absolute size. We show that when the temperature is varied, the growth and division timescales scale proportionally with each other over the physiological temperature range. 
%The temperature dependence of the mean growth rate obeys an Arrhenius Law, with an activation energy that is typical of an enzyme catalyzed reaction. 
Strikingly, the  cell-size and division-time distributions can both be rescaled by their mean values such that the condition-specific distributions collapse to universal curves. We account for these observations with a minimal stochastic model that is based on an autocatalytic cycle.  It predicts the scalings, as well as specific functional forms for the universal curves. Our experimental and theoretical analysis reveals a simple physical principle governing these complex biological processes:  a single temperature-dependent scale of cellular time governs the stochastic dynamics of growth and division in balanced growth conditions. 
\end{abstract}
%\sib{[The last clause is necessary to clarify what the physical insight gained is.]}

\maketitle

\section{Significance}
Growth and division of individual cells are the fundamental events underlying many biological processes, including the development of organisms, the growth of tumors, and pathogen-host interactions. Quantitative studies of bacteria can provide insights into single-cell growth and division but are challenging owing to the intrinsic noise in these processes.   Now, by using a unique combination of measurement and analysis technologies, together with mathematical modeling, we discover quantitative features that are conserved across physiological conditions.   These universal behaviors reflect the physical principle that a single timescale governs noisy bacterial growth and division despite the complexity of underlying molecular mechanisms. 

\section{Introduction}
Quantitative studies of bacterial growth and division initiated the  molecular biology revolution~\cite{1949-monod-uq} and continue to provide constraints on molecular mechanisms \cite{1949-monod-uq, 1958-schaechter-ly,1968-donachie-mi,2010-scott-vn, 2009-tzur-dq,2011-huh-zr,2012-lin-uq,  2010-hagen-fk}. Yet many basic questions about the growth law, i.e., the time evolution of the size of an individual cell, remain \cite{2011-scott-kx, 2012-velenich-vn, 2008-haeusser-bs, 2010-hagen-fk, 2013-mcmeekin-nx, 2009-halter-uq}.  Whether cells specifically sense size, time, or particular molecular features to  initiate cell division is also unknown \cite{2007-talia-tg}.
Answers to these questions, for individual cells {in balanced growth conditions}, are of fundamental importance, and they serve as starting points for understanding collective behaviors involving {spatiotemporal interactions between} many cells \cite{Goldenfeld2007,Hawkins2007,Zhang2012,Idema2013}.

Cell numbers increase exponentially in bulk culture in balanced growth conditions irrespective of how the size of an individual cell increases with time \cite{1949-monod-uq}. Thus observation of the population is insufficient to reveal the functional form of the growth law for a given condition. Bulk culture measurements necessarily average over large numbers of cells, which can conceal cell-to-cell variability in division times, sizes at division, growth rates, and other properties \cite{2007-maheshri-bh}. Moreover, the cell cycles of different cells in the population are typically at different stages of completion at a given time of observation.   Even when effort is made to synchronize cells at the start of an experiment, so as to have a more tightly regulated initial distribution of growth phases, this dispersion can only be mitigated, not eliminated.  These considerations highlight the importance of studying growth and division at the single-cell level.
 
The landmark papers of Koch, Neidhardt, Schaetchter, and co-workers~\cite{1962-schaechter-dz, 1962-koch-kl, 1958-schaechter-ly} addressed  issues of growth at the single-cell level, but the (statistical) precision of these measurements was not sufficient to characterize the growth law(s) under different conditions. There is evidence that the growth law under favorable conditions is generally exponential  \cite{Elliott1978, Cooper1988, 2007-talia-tg, 2010-godin-uq, 2010-wang-cr}. However, both linear and exponential growth laws have been previously proposed \cite{1991-cooper-sf,2001-koch-yg,2006-cooper-rm,1968-kubitschek-lq}.  Furthermore, it is estimated that a measurement precision of 6\% is required to discriminate between these functional forms for cells that double in size during each division period \cite{2009-tzur-dq}. This precision is difficult to achieve in typical single-cell microscopy studies because cell division leads to rapid crowding of the field of view \cite{2013-ullman}.  

{Various experimental approaches have been introduced to address this issue\cite{1963-helmstetter,2012-manalis-babymachine,2013-ouyang-babymachine, 2012-moffitt, 2010-wang-cr}.  Conventional single-cell measurements on agarose pads are limited to about 10 generations, and the age distribution of the observed cells is skewed towards younger cells because the population numbers grow geometrically \cite{2009-locke}. Designed confinement of cells allows observation of constant numbers of cells without requiring genetic manipulation \cite{2012-moffitt, 2010-wang-cr}.  The system that we describe here for {\it Caulobacter crescentus} also allows tracking constant numbers of single cells over many generations at constant (and, if desired, low) number densities.  This setup provides the advantages that contacts between cells can be avoided and the environment can be kept invariant over the course of an experiment, such that all cells exhibit equivalent statistics. In fact, in control experiments with this system, we observe that cells grow at reduced rates when they come in contact with each other. Our extensive data provide the statistical precision needed to transcend previous studies in order to establish the functional form of the mean growth law under different conditions and to characterize fluctuations in growth and division.}

\section{Results and Discussion}

\subsection{Experimental design}

Determining quantitative laws governing growth and division requires precise measurement of cell sizes {of growing cells} under invariant conditions for many generations.  We achieved these criteria by choosing an organism that permits control of cell density through molecular biology and microfluidics.  The bacterium {\it C.\ crescentus} divides into two morphologically and functionally distinct daughter cells: a motile swarmer cell and an adherent stalked cell that is replication competent. A key improvement over our earlier work~\cite{2012-lin-uq,2010-lin-fk} is that the surface adhesion phenotype can be switched on/off with an inducible promoter.  This strain,  in combination with automated microscopy in a temperature-controlled enclosure, allows measurement of $\sim$1000 single stalked cells for $>$100 generations each at constant low density (uncrowded) balanced growth conditions (SI Text Section 1 and Fig.\  S1). 

%\sib{[Should we explicitly mention here that we are only tracking one portion of the life cycle of CC?]}

We determine the area of {\em each} stalked cell in our two-dimensional images with a precision better than 2\% (Methods, SI Text Section 2, and Figs.\  S1 and S2).  Since {these} cells are cylindrically symmetric  around the curved longitudinal axis, the measured areas account for the varying width of the cell and faithfully report the cell volumes (Fig.\  S3).  We thus use cell areas to quantify cell sizes.  Using image processing software that we developed, we obtain 4,000 to 16,000 growth curves for individual cells in complex medium (PYE) at each of seven temperatures spanning the physiological range of the organism: 14, 17, 24, 28, 31, 34, and 37$^{\circ}$C. 

\subsection{Cells sizes increase exponentially to a relative threshold; \\mean growth rate determines mean division time}

\begin{figure}[h]
\begin{center}
\includegraphics[width=0.5\columnwidth]{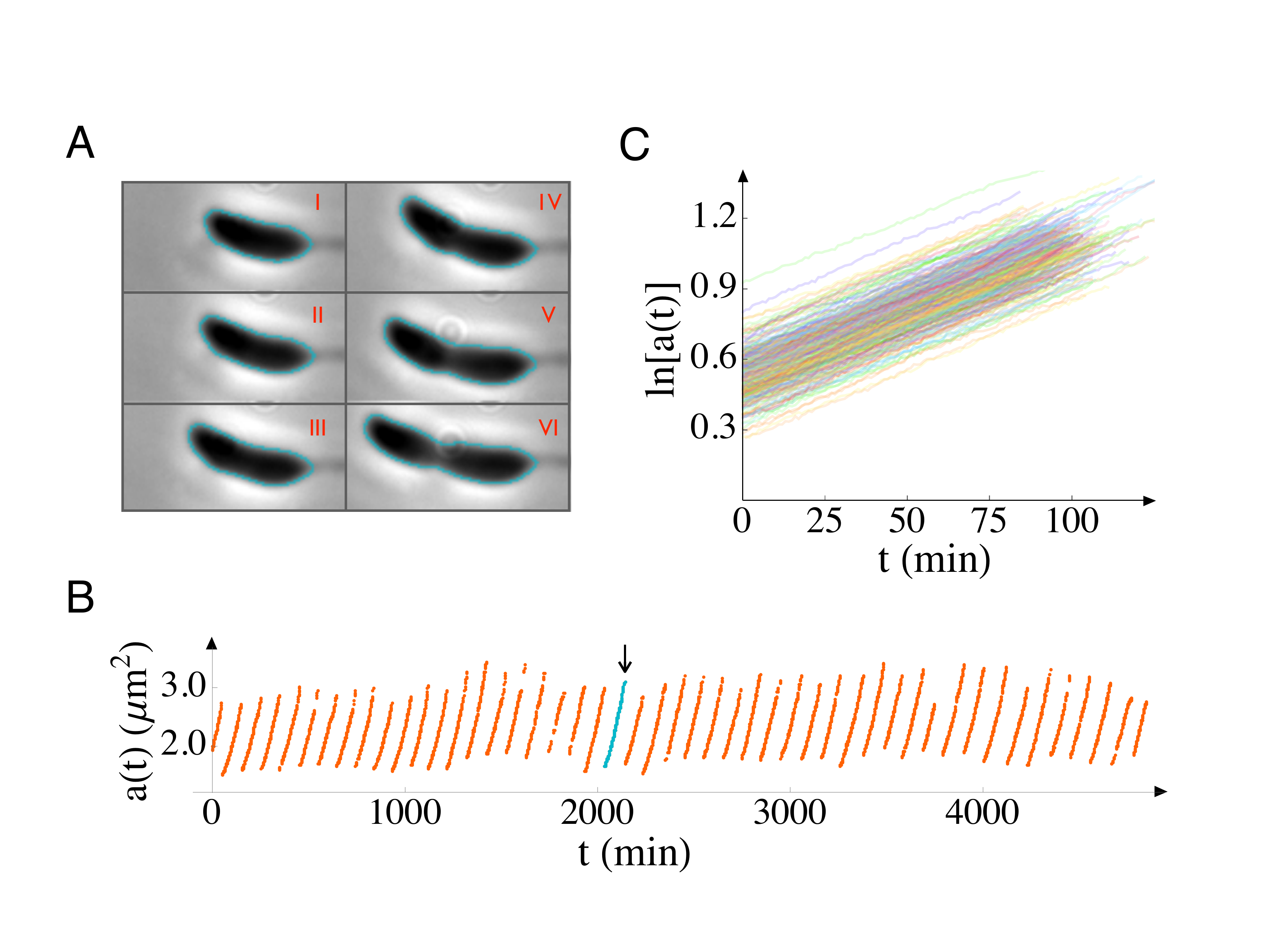}
\caption{ {\bf Cell sizes (areas, $a$) as functions of time ($t$).} 
{{\bf (A)} Six phase contrast images of a cell, all taken from a single generation at 15 min intervals, starting from 10 min after the previous division, are shown (respectively labelled I-VI).} From such images, the area of each cell as a function of time is inferred from the outlines indicated. {\bf (B)} The area is plotted as a function of time for {many generations of a single cell. The generation indicated in teal and by an arrow indicates time period from which images in (A) are taken.} {\bf (C)} We plot measured areas on semi-logarithmic scale to make the growth law evident. The data shown are from 5 cells over the course of $\approx$300 generations each in an experiment conducted at 24$^\circ$C in complex medium. More than 4,000 growth curves were obtained from $\approx$100 cells in this experiment; only a small subset are shown here for clarity. The image acquisition rate is 1 frame/min.}
\label{fig-exp-growth}
\end{center}
\end{figure}

\begin{figure}[ht]
\begin{center}
\includegraphics[width=0.5\columnwidth]{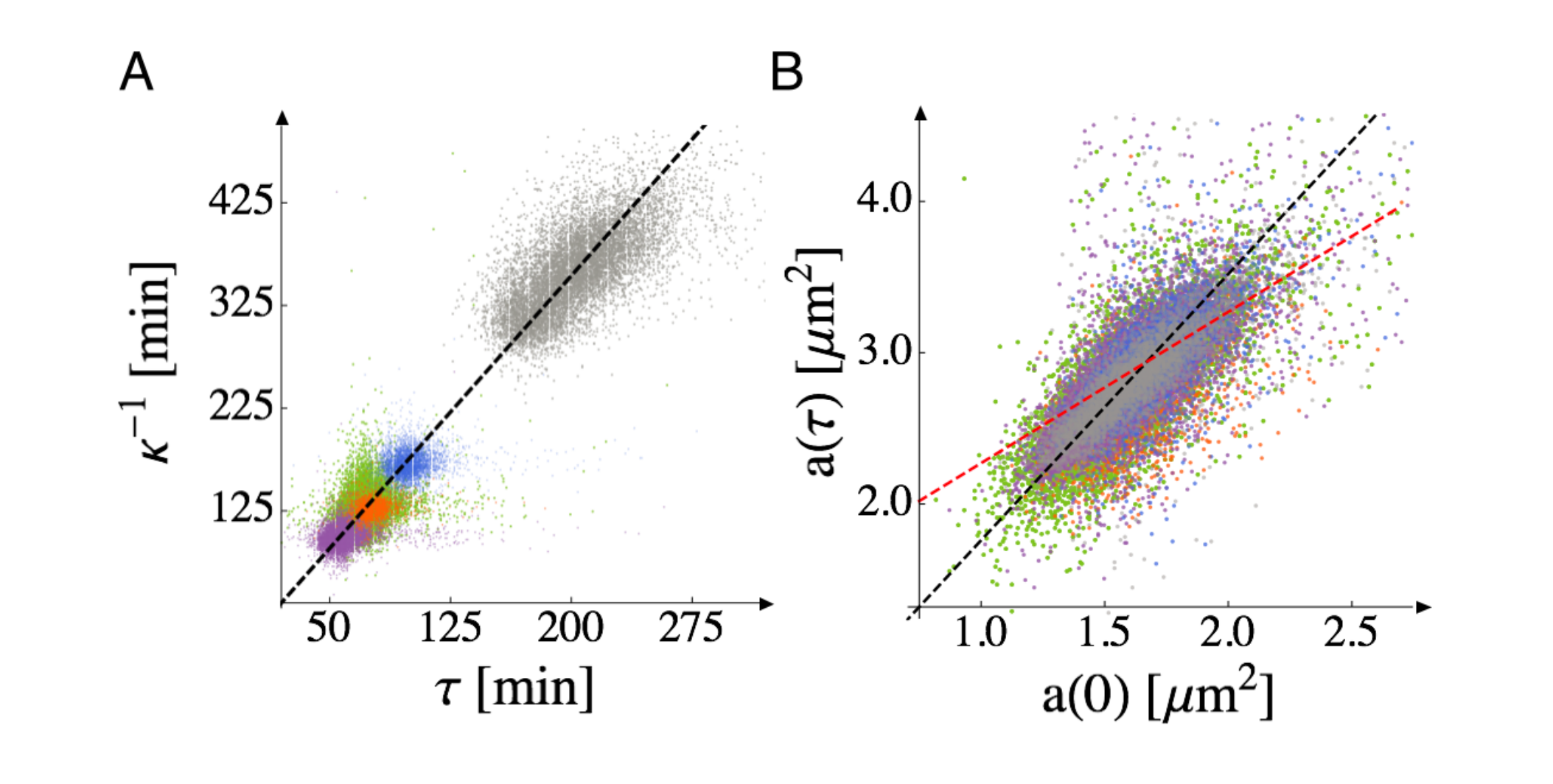}
%\vspace*{-2cm}
\caption{ {\bf  Proportionality of the growth and division timescales; cell size at division is a critical multiple of its initial size.} 
Superposition of data from temperatures across the physiologically relevant range (purple, $34^{\circ}$C; green, $31^{\circ}$C; orange, $28^{\circ}$C; blue, $24^{\circ}$C; gray, $17^{\circ}$C).  There are  4,000-16,000 data points for each temperature. 
{\bf (A)}  Points are obtained by identifying division periods $\t_{ij}$ and fitting single-cell growth trajectories to the exponential growth law, Eq.\ \ref{eq-exp}. The slope of the best fit line (shown in black) is $0.565$, which is equivalent to division occurring when $\la a(\t;T)/a(0;T) \ra \approx \exp{(0.565)} = 1.76$. The coefficient of determination for the fit is $R^{2} = 0.98$ for all temperatures.   (The faint banding is a visualization artifact rather than a feature of the data.)
 {\bf (B)}  The final area just prior to division, $a_{ij}(\t;T)$, is plotted against the initial area, $a_{ij}(0)$, of each cell. The data from all five temperatures are scattered around the black dashed straight line $a(\t;T) =  1.76 \, a(0)$. $R^{2} = 0.99$ for all temperatures.  The red dashed line represents division at constant swarmer cell size for comparison.} 
\label{fig-beta-tau}
\end{center}
\end{figure}

Fig.\  \ref{fig-exp-growth} shows representative data for single-cell growth.  The fact that the curves are straight on a semi-logarithmic plot indicates that the growth law is exponential (see also Fig.\ S4); this relation holds for all temperatures studied.  In other words, each growth curve can be well fit by the form 
%%%%%%%%%%%%%%%%%%%%%%%%begin%%%%%%%%%%%%%%%%%%%%%%%%
\begin{align}\label{eq-exp}
a_{i j}(t;T) = a_{i j}(0;T) \exp[\k_{i j}(T)t],
\end{align}
%%%%%%%%%%%%%%%%%%%%%%%%%end%%%%%%%%%%%%%%%%%%%%%%%
where $a_{i j} (0;T)$ is the initial size of the $i^{\mbox{\tiny{th}}}$ stalked cell in the $j^{\mbox{\tiny{th}}}$ generation, and $T$ is the temperature.  Each growth curve yields a division time, $\t_{i j}(T)$, and a rate of exponential growth, $\kappa_{i j}(T)$ (Fig.\ \ref{fig-beta-tau} and SI Text Section 3.1).  

Fig.\  \ref{fig-beta-tau} shows the parameters in Eq.\ \ref{eq-exp} for each growth curve at each temperature.  The growth and division time scales,  $\k_{ij}^{-1}(T)$ and $\t_{ij}(T)$, respectively, vary proportionally (over about a four-fold dynamic range; Fig.\  \ref{fig-beta-tau}A), such that the mean growth rate and mean division time determine each other.  This fact, together with Eq.\ \ref{eq-exp}, suggests that the initial and final sizes of the cells should also scale linearly with each other (with no additive offset), to be consistent with  exponential growth.  We confirm experimentally that they do (Fig.\  \ref{fig-beta-tau}B), which further supports the exponential growth law (see also SI Text Section 3.2 and Figs.\ S5 and S6).  

The biological significance of Fig.\ \ref{fig-beta-tau}B is that cells divide when their sizes are a constant multiple of the initial stalked cell size.  The existence of a  relative size threshold is further supported by the fact that the ratio $a_{ij}(\t;T)/a_{ij}(0;T)$ appears more tightly regulated than $a_{ij}(\t;T)$ (see Fig.\ S7) as their respective coefficients of variation (standard deviation divided by mean) are $\approx$8\% and $\approx$20\%.  From the slope of the best fit line in black in Fig.\  \ref{fig-beta-tau}B, we obtain $\la a_{ij}(\t;T)/a_{ij}(0;T) \ra\approx \exp(0.565) = 1.76$, where $\la ... \ra$ indicates a population average. This value is consistent with known {\it average} ratios of stalked and swarmer cell sizes for {\it C.\ crescentus} \cite{1964-poindexter-fk}, but prior measurements could not eliminate alternative single-cell scenarios.  For example, one might have just as well have expected division at constant swarmer cell size, in analogy to budding yeast \cite{2007-talia-tg} or the model proposed in [38] for symmetrically dividing bacteria; in that case, the points would follow a line with a slope of 1 and a non-zero intercept, as indicated by the red dashed line in Fig.\ \ref{fig-beta-tau}B.  An important implication of the relative size threshold is that there must be growth during the swarmer stage; whether this growth occurs throughout the swarmer stage or together with differentiation remains to be demonstrated.

\subsection{Mean division time decreases as temperature increases}
%quantitative comparison is not appropriate because we observe only stalked cells, whereas bulk cultures also include swarmer cells, and the temperature dependence of the swarmer-to-stalked transition time is not known.  

\begin{figure}[ht]
\begin{center}
\includegraphics[width=0.5\columnwidth]{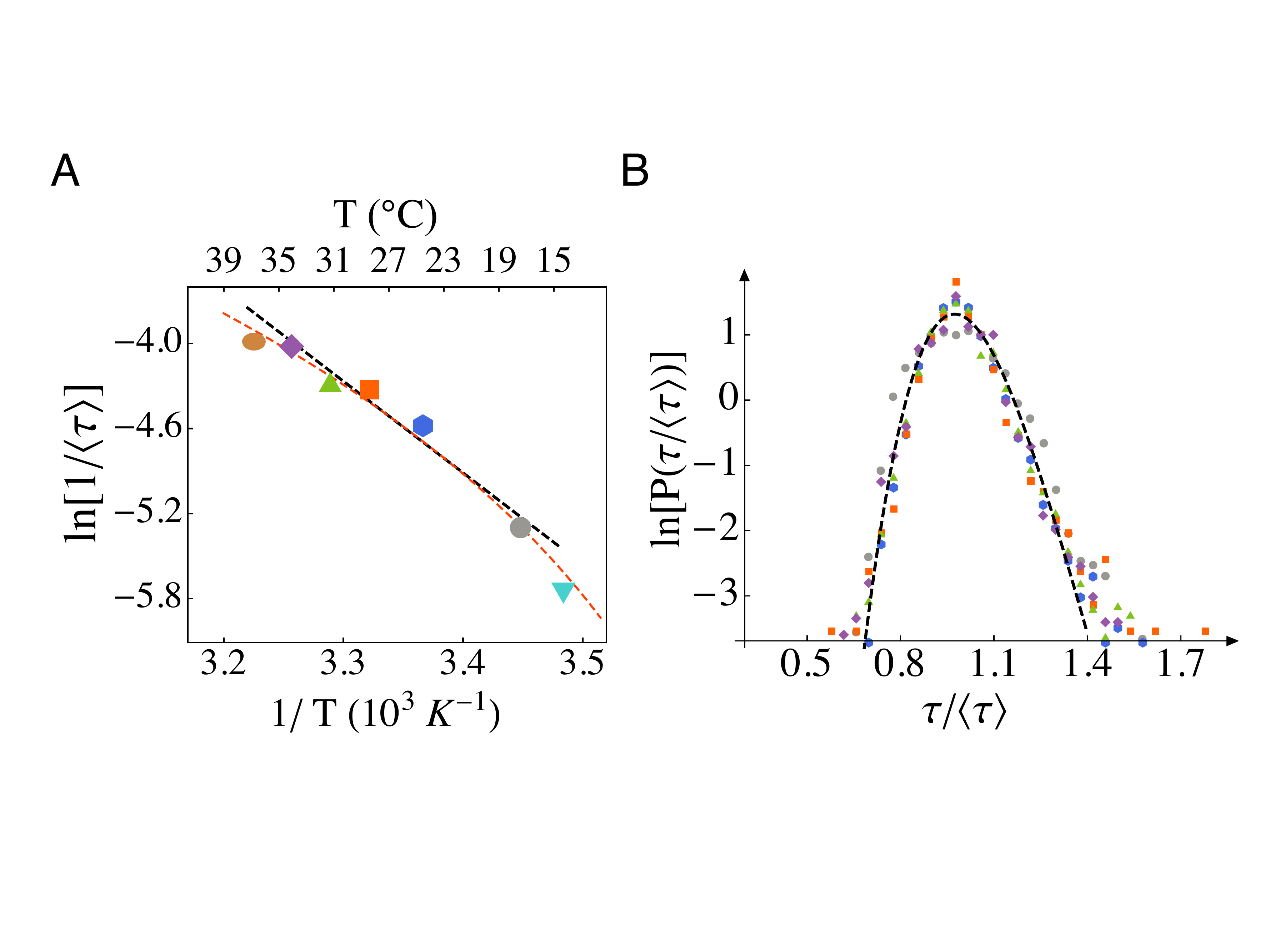}
%\vspace*{-2cm}
\caption{{ {\bf Scaling of the division time distribution with temperature.} {\bf (A)} Variation of the mean division time with temperature ({brown, $37^\circ$C}, purple, $34^{\circ}$C; green, $31^{\circ}$C; orange, $28^{\circ}$C; blue, $24^{\circ}$C; gray, $17^{\circ}$C, {cyan, $14^\circ$C}; ``ln'' is natural logarithm); the error in the mean is less than the size of the symbols. The effective activation barrier is inferred to be $\Delta E = 54.0$ kJ/mol (12.9 kcal/mol)  from the slope of the fit to the data over the temperature range that the data approximately follows an Arrhenius form, 17-34$^\circ$C ($R^{2} =  0.97$). The red dashed line is a fit of the Ratkowsky form, $\la\t\ra^{-1}\sim (T - T_0)^2$ \cite{1982-ratkowsky-xy}, over the entire temperature range studied; $T_0$ is inferred to be 270 K.  We also provide a Celsius scale (top) for convenience; note that this scale is not linear.  {\bf (B)}  Probability distributions of division times from different temperatures  (colors are the same as in (A)), rescaled by the respective temperature-dependent mean values in (A), collapse to a single curve (COV $\approx$13\%).  The invariant shape of the distribution indicates that  a single timescale, expressible in terms of the mean division time, governs stochastic division dynamics.}
}
\label{fig-ig}
\end{center}
\end{figure}

We plot the logarithm of the growth rate against the inverse temperature, as is common for bulk culture studies \cite{1983-ingraham-qf, 1983-ratkowsky-sf, 1982-ratkowsky-xy}, in Fig.\ \ref{fig-ig}A.  For bulk culture studies, such plots typically deviate from a strict Arrhenius Law (a straight line in Fig.\ \ref{fig-ig}A, corresponding to $\la \k\ra \propto \la \t(T) \ra^{-1} = A\exp[-\Delta E/k_BT]$,  where $A$ is a temperature-independent constant, $\Delta E$ is the activation energy, and $k_{B}$ is Boltzmann's constant) \cite{1983-ratkowsky-sf, 1982-ratkowsky-xy} and exhibit a turnover in the growth rate.  We do not observe a turnover in the single-cell growth rate over the temperatures studied, which span the physiological range---the mean division time decreases as the temperature increases over the full range (although see {\it Extreme temperatures reinforce the scaling laws} for a discussion of mortality).  

{The points in the range 17 to 34$^\circ$C fall sufficiently near a straight line that one can use the data to estimate an effective $\Delta E$, also known as the ``temperature characteristic'' \cite{1983-ingraham-qf, 1983-ratkowsky-sf, 1982-ratkowsky-xy}.  We find $\Delta E = 54.0$ kJ/mol  (12.9 kcal/mol), which is consistent with previous estimates from bulk culture measurements for several bacteria \cite{1942-monod-kx,1979-herendeen-qf,1971-shaw}.  Empirical relations have been proposed to capture the negative curvature in Fig.\ \ref{fig-ig}A, and we show that the best fit of the form suggested by Ratkowsky and co-workers \cite{1983-ingraham-qf, 1983-ratkowsky-sf, 1982-ratkowsky-xy}, $\la\t\ra^{-1}\sim (T-T_0)^2$, in Fig.\ \ref{fig-ig}A.  In that model, the ``minimum temperature'' $T_0$ sets the energy scale; for our data, $T_0= 270$ K.  A series expansion shows that values in the range 260-280 K, as tabulated for other microorganisms \cite{1983-ratkowsky-sf, 1982-ratkowsky-xy}, are consistent with $\Delta E\approx 54$ kJ/mol (see SI Text Section 4.1 for further discussion).  The precise values of parameters of course depend on the temperature ranges used for the fits, but it is important to note that $\Delta E$ is of the order a typical enzyme-catalyzed reaction's activation energy \cite{1943-sizer, Abbondanzieri2005}.

%We focus here on the Arrhenius regime and discuss the extreme temperatures (14 and 37$^{\circ}$C) below. 

%Since Fig.\ \ref{fig-beta-tau}A demonstrates that the mean growth rate is proportional to the inverse of the mean division time, the mean growth rate also follows an approximate Arrhenius Law, with the same activation energy. 

%This mechanism contrasts with one involving a single rate-limiting step---such a mechanism would require most of the steps to have unusually low barriers.  

\subsection{Model for exponential growth}
Motivated by our observations for the mean behaviors,} we consider a simple kinetic model that was introduced by Hinshelwood to describe exponential growth in 1952 \cite{1952-hinshelwood}. This model consists of an autocatalytic cycle of $N$ reactions, in which each species catalyzes production of the next (Fig.\ \ref{fig-hinshelwood}).   An important feature of this model is that the overall rate constant for growth ($\kappa$) is the geometric mean of the rate constants of the elementary steps ($k_i$) \cite{1952-hinshelwood} (see Fig.\ S8):
\begin{equation} \label{eqn-geom}
\kappa = (k_1k_2\ldots k_N)^{1/N}.
\end{equation} 
Therefore, if the rates of the elementary steps vary in an Arrhenius fashion,  the overall rate constant for growth must  vary similarly.  To see this, substitute $k_i(T)= A_i\exp[-\Delta E_i/k_BT]$ (where $A_i$ and $\Delta E_i$ are the collision frequency and activation energy of reaction $i$) into Eq.\ \ref{eqn-geom}:
\begin{eqnarray}
\kappa(T) &=& (A_1\ldots A_N)^{1/N}\exp\left[-\frac{\Delta E_1 + \ldots + \Delta E_N}
{Nk_BT}\right]\\ \nonumber 
& \equiv& A\exp[-\Delta E/k_BT].
\end{eqnarray}
This equation shows that $\Delta E$ is the arithmetic mean of the elementary activation energies. Therefore, if each step has an activation energy of the order of a typical enzyme reaction's, then so does the effective growth rate. This idea is consistent with our measurements (Fig.\  \ref{fig-ig}A), and is independent of the chemical identities of $X_{i}$ and the value of $N$. 

{It is important to stress that the validity of the model and its conclusions are not contingent on a specific form for the temperature dependence.  While it is arguably easiest to see the averaging of the rate in the Arrhenius case considered above, it is generally true that the overall rate varies like the constituent rates. For example, if the constituent rates follow the Ratkowsky form \cite{1982-ratkowsky-xy}, then the composite rate does as well, to leading order, so long as the energy scales of the individual steps are not very disparate (see SI Text Section 4.1).}

\subsection{Fluctuations in cell sizes scale with their means}

%For the case $N = 2$, a molecular interpretation of the model is that one species represents the protein synthesis machinery (ribosomes and associated proteins) and the other represents proteins for metabolism \cite{1988-koch}, but it could represent other cell processes as well.  An example of In particular, the measured activation energy for transcription is $\approx$13 kcal/mol \cite{Abbondanzieri2005}. This value makes it reasonable for transcription to be a step in the cycle, but this does not preclude other reactions with comparable barriers from also contributing since $\Delta E$ is an average.  

Given that the Hinshelwood cycle captures the mean behaviors that we observe, it is of interest to understand its implications for the fluctuation statistics that we can obtain from our extensive single-cell growth data. To this end, we recently generalized the model by assuming that the reactions in the cycle have exponential waiting-time distributions and showed analytically that its dynamics reduces to those of a single composite stochastic variable \cite{iyer-biswas-fk}.  We term this model the stochastic Hinshelwood cycle (SHC). 
%First, the mean behavior of the SHC recapitulates that of the original deterministic one. Specifically, exponential growth with an effective growth rate, $\k$, equal to the geometric mean of all the individual rates results. 

\begin{figure}[ht]
\begin{center}
\includegraphics[width=0.4\columnwidth]{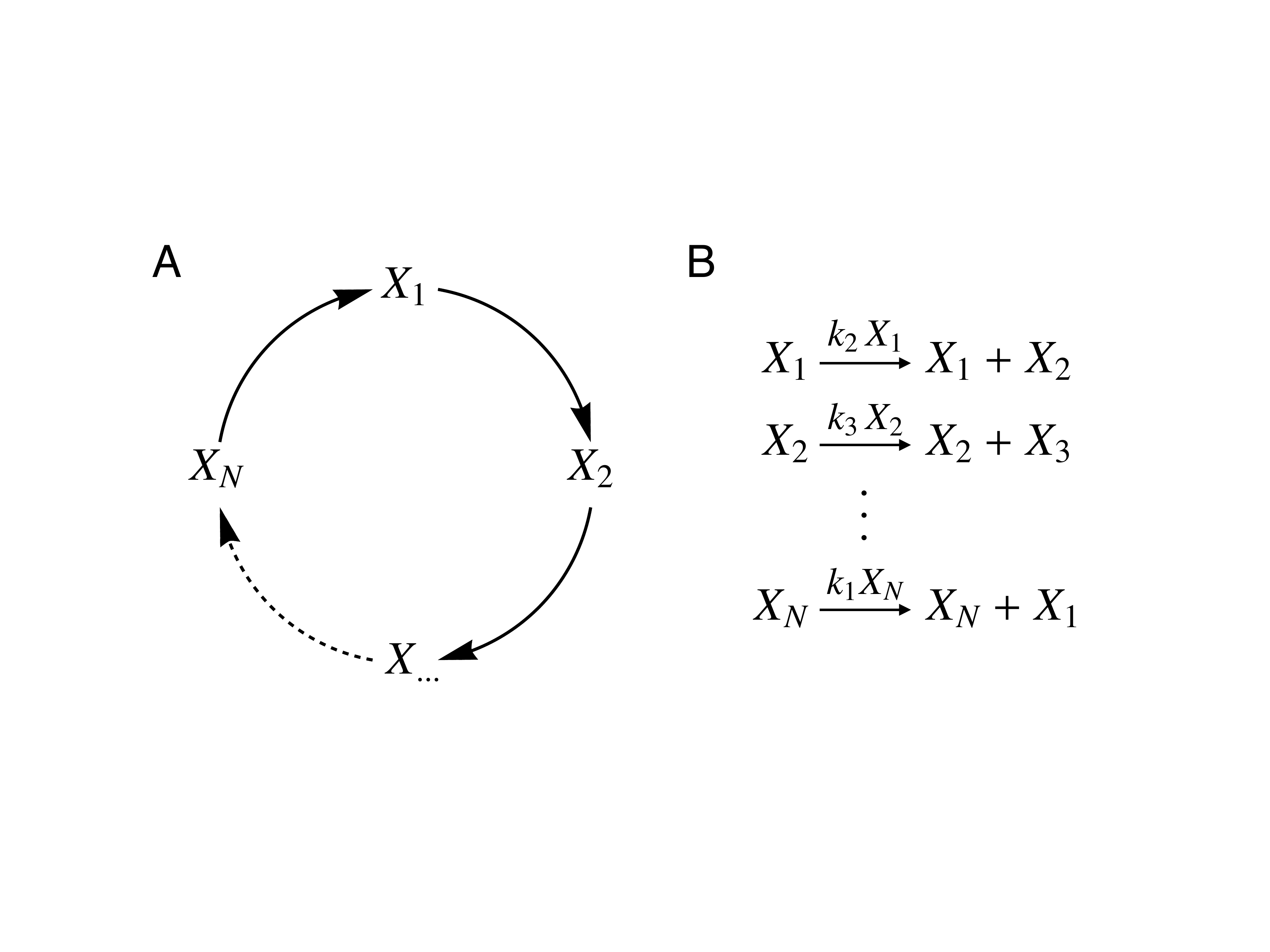}
\caption{{\bf Hinshelwood model for exponential growth.} {\bf (A)} Schematic showing the autocatalytic cycle, in which each species activates production of the next.  {\bf (B)} Corresponding reactions. The size of a cell is assumed to be proportional to a linear combination of the copy numbers of the species in the cycle.  {In the stochastic Hinshelwood cycle (SHC), the dwell times are assumed to be exponentially distributed; reaction propensities are indicated above the arrows in (B).  Note that the effective growth rate ($\kappa$) depends only on the rate constants ($k_i$).}
}
\label{fig-hinshelwood}
\end{center}
\end{figure}

\begin{figure}[ht]
\begin{center}
\includegraphics[width=0.5\columnwidth]{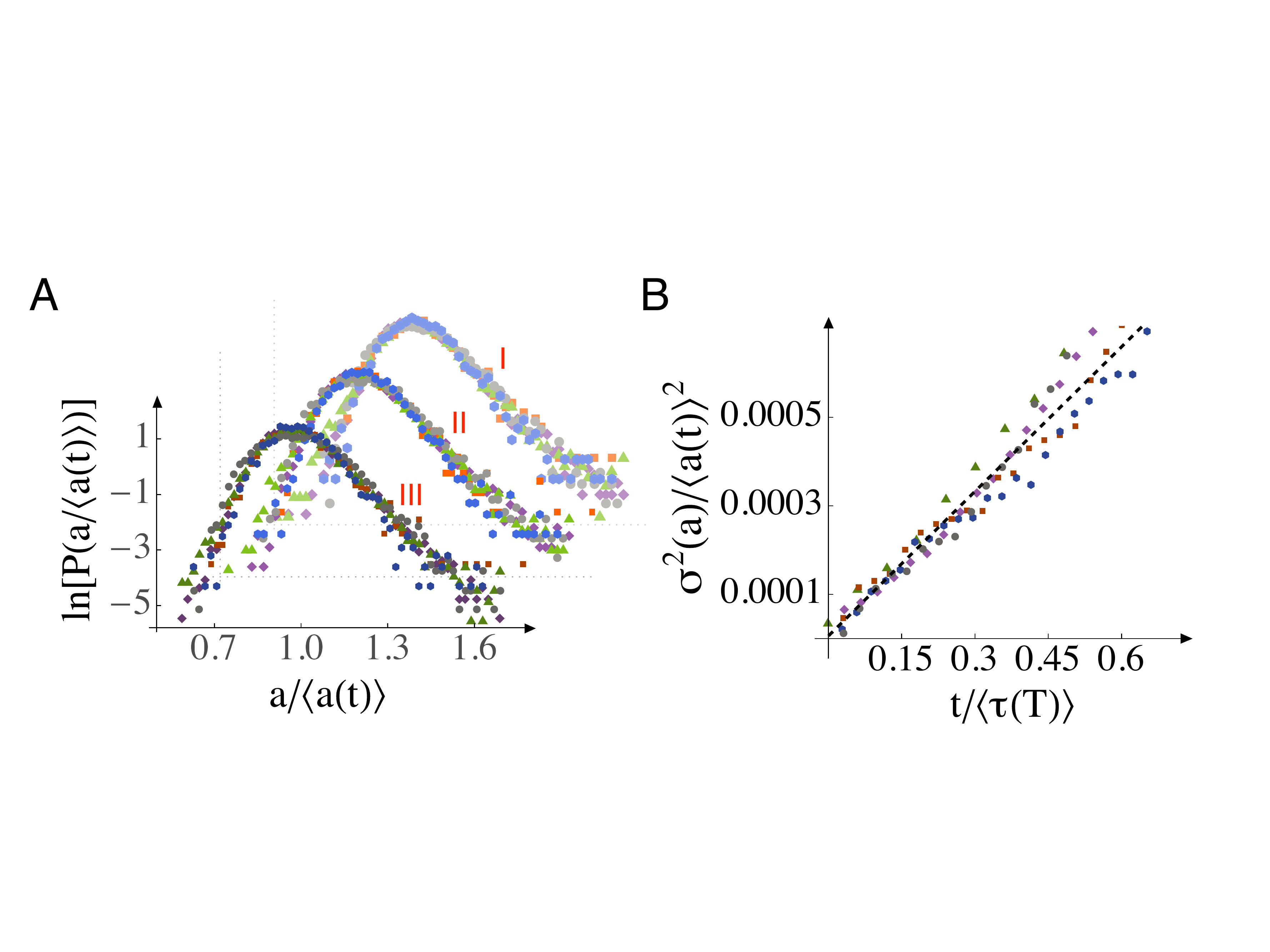}
\vspace*{-1cm}
\caption{  {\bf Scaling of cell-size fluctuations within each division period.} {\bf (A)} The size (area) distributions at all temperatures (purple, $34^{\circ}$C; green, $31^{\circ}$C; orange, $28^{\circ}$C; blue, $24^{\circ}$C; gray, $17^{\circ}$C) are plotted for three different rescaled time points, at $t/\la \t(T) \ra = 0, 0.2$, and 0.6 (marked I, II, and III, respectively). ``ln'' is natural logarithm. The area distributions at each time have been rescaled by their exponentially growing mean sizes.  {\bf (B)}  Relaxation of the coefficient of variation of cell size (area) after division.  The slope of the black dashed line, which is fitted to data for all temperatures simultaneously, is 0.0011.}
\label{fig-sizedist}
\end{center}
\end{figure}

A key result of the model is that, asymptotically, fluctuations in all chemicals in the SHC (Fig.\ \ref{fig-hinshelwood}) become perfectly correlated with each other~\cite{iyer-biswas-fk}. Thus, the SHC makes a strong prediction for the scaling of size fluctuations in the asymptotic limit:  the cell-size distributions from all times should collapse to a universal curve when they are rescaled by their exponentially growing means, \cite{iyer-biswas-fk}. In other words, in balanced growth conditions, the width of the size distribution grows exponentially at the same rate as the mean growth rate, $\k$. This prediction is validated by our data in the Arrhenius range, as shown in Fig.\  \ref{fig-sizedist}A (see also SI Text Section 4.2).

The SHC also makes predictions for dynamics of growth noise in {\em individual} growth trajectories.  To enable comparison to our data, we have derived the equivalent Langevin description from the Master equation for the SHC. It is \cite{iyer-biswas-fk}
%%%%%%%%%%%%%%%%%%%%%%begin%%%%%%%%%%%%%%%%%%%%%%
\begin{align}\label{eq-le}
\frac{ d\, a(t;T)}{d\, t} = \k(T) \, a(t;T) + \eta(t;T)\sqrt{ a(t;T) }, 
\end{align}
%%%%%%%%%%%%%%%%%%%%%%%end%%%%%%%%%%%%%%%%%%%%%%
where $\eta$ is Gaussian white noise satisfying $\la \eta(t_{1};T)\eta(t_{2},T)\ra \equiv B(T)\delta({t_{1}-t_{2}})$, which defines $B(T)$~\cite{iyer-biswas-fk}.  The first term on the right hand side of Eq.\ \ref{eq-le} represents the systematic exponential growth (the ``drift'') and the second term is the noise (the ``diffusion'').  Eq.\ \ref{eq-le} shows that the noise increases in magnitude with the area (i.e., it is ``multiplicative''), and it does so in proportion to the square root of the area.  This Langevin equation contrasts with the well-known Black-Scholes equation for multiplicative noise (also known as geometric Brownian motion), {which has been invoked to explain cell size distributions \cite{2005-furusawa-rr}. In the Black-Scholes equation,} both the drift and diffusion terms scale linearly with the dynamical variable. The square root multiplicative noise in Eq.\ \ref{eq-le} results in the observed scale invariance of the cell size distribution, and the corresponding mean rescaled asymptotic cell size distribution is a gamma distribution~\cite{iyer-biswas-fk}. In contrast, the Black-Scholes equation does not yield the observed  constancy of the coefficient of variation of cell sizes, and instead predicts a lognormal cell-size distribution with a coefficient of variation that increases as $\sqrt{t}$. 

For a given initial cell size, Eq.\ \ref{eq-le} predicts that the square of the coefficient of variation of cell sizes should fall on a straight line when plotted against time; additionally, dimensional analysis dictates that the slope of this straight line should be independent of temperature when we rescale $B(T)$ by $\k(T)$ and $t$ by $\la \t(T) \ra$:
%%%%%%%%%%%%%%%%%%%%%%begin%%%%%%%%%%%%%%%%%%%%%%
\begin{equation}
\label{eq-sc-td}
 \frac{\s^{2}(t;T)}{\la a(t;T) \ra^{2}} \approx \frac{{B(T)t}}{a(0;T)} \approx \frac{1}{a(0;T)} \left[\frac{B(T)}{\la\k(T)\ra}\right] \frac{t}{{\la\t(T)\ra}}.
\end{equation}
%%%%%%%%%%%%%%%%%%%%%%%end%%%%%%%%%%%%%%%%%%%%%%
In Fig.\  \ref{fig-sizedist}B we show that this prediction is also validated by our data, and that $B(T)/\la \k(T)\ra= 0.0011$.  This value indicates that the fluctuations around each individual exponential growth curve are small compared with its time constant.  To the best of our knowledge, the SHC is the only microscopic model of stochastic exponential growth to capture the statistics of individual growth trajectories that we measure (Figs.\ \ref{fig-sizedist}A and \ref{fig-sizedist}B).

\subsection{Fluctuations in division times scale with their means}

Next, we examine fluctuations in cell division times and their variation with temperature. We show in [48] that treating stochastic division of cells as a first passage time problem for the cell size to reach a critical value gives rise to additional scaling forms. The mean-rescaled division time distributions from all temperatures should collapse to the same curve, since the single timescale, $\la \k (T) \ra^{-1}$, which is proportional to $\la \t(T) \ra$ (Fig.~\ref{fig-beta-tau}), governs stochastic division dynamics\cite{iyer-biswas-fk}. This prediction is validated by our observed division time distributions from all temperatures (Fig.\ \ref{fig-ig}B). Specifically, the SHC model predicts a beta-exponential distribution of division times for an absolute cell-size threshold and a given initial size \cite{iyer-biswas-fk}. By convolving this result with the observed initial size distribution (Fig.~\ref{fig-sizedist}), we can determine the expression for the division time distribution for the observed relative size thresholding (Fig~\ref{fig-beta-tau}B; see Supplementary Section 3 for details).  This form provides a good fit of the data (Fig.\  \ref{fig-ig}B and SI Text Section 4.3) for all temperatures in the Arrhenius range (17-34$^\circ$C). 
%\cite{iyer-biswas-fk} 
\subsection{Extreme temperatures reinforce the scaling laws}

Finally, we discuss the behavior outside of the Arrhenius range. At $37^{\circ}$C, there is significant cell mortality.  The probability of a cell surviving is a decaying exponential function of time, corresponding to a constant probability per unit time of dying of 
$\approx$7 \% per mean cell lifetime (see SI Text Section 5 and Fig.\ S9 for details). At all other temperatures (in PYE medium), cell mortality is less than 1\% for up to a 100 generations, and we do not observe any senescence (i.e., systematic decrease in reproductive output with time) \cite{2006-fredriksson-uq}. Bulk-culture measurements cannot separate the contributions to decreased reproductive output from increased mortality and decreased  growth rates of surviving cells. 

%(we did not observe division within $24$ h at $10^{\circ}$C, although the cells were not not killed and resumed growth and division when warmed).  

Remarkably, at both 14$^\circ$C and 37$^\circ$C, the single-cell growth law for surviving cells continues to be exponential (Fig.\ S10), and the exponential growth timescale,  $\la \k(T) \ra^{-1}$, continues to scale proportionally with the mean division time (Fig.\ S11).  In other words, both growth and division slow together. Consequently, the final size at division also scales proportionally with the initial size of the cell and thus a relative cell size thresholding scheme for cell division continues to hold at these temperatures.  The scaling laws for mean-rescaled cell size and division time distributions also continue to hold for these two temperatures (SI Text Section 5 and Fig.\ S11).  Our results for growth at extreme temperatures further validate the scaling predictions and show that they continue to hold even when the growth rate deviates from the Arrhenius Law.

\subsection{Applicability to other microorganisms}
While the results presented here are for {\it C.\ crescentus} in complex medium, we expect them to apply to growth and division of other microorganisms in different balanced growth conditions.  In [48], we show that the size scaling laws follow directly from exponential growth, and, as noted in the Introduction, there is evidence of exponential growth in several bacteria, including {\it E.\ coli}  \cite{Elliott1978, Cooper1988, 2007-talia-tg, 2010-godin-uq, 2010-wang-cr}.  We thus expect the cell size distributions of these organisms to collapse to a single curve when rescaled by their means.
%\cite{iyer-biswas-fk}
% [ARD:  I commented the next sentence out since it seems too speculative given the fact that these are bulk measurements.]
%Indirect evidence for the validity of the SHC for other exponentially growing cells is provided by the observation that the activation energy barrier (temperature characteristic), in the Arrhenius range, is $\approx$13 kcal/mol in bulk culture measurements for several bacteria ~\cite{1983-ingraham-qf, iyer-biswas-fk}. The Ratkowsky minimum temperature for our data is  $T_0 \approx$ 269K~\cite{1982-ratkowsky-xy, 1983-ratkowsky-sf}. $T_0$ in the range of $270-280$K, as tabulated by Ratkowsky et-al for several bacteria, also corresponds to an activation energy of $\approx$ 13 kcal/mol in the Arrhenius range for all of them. [This can be seen by series expansion of Ratkowsky's empirical relation around $300$K~\cite{1982-ratkowsky-xy}.]
The premise that size scaling generally holds for bacteria in balanced growth conditions was put forth long ago \cite{1982-trueba-ve}; however, it is important to note that the size distribution in earlier studies was a convolution of our size distribution with the cell-cycle-phase distribution (related to our division-time distribution) because the data were taken from images at single laboratory times for asynchronous populations.  

%It is interesting to consider whether the relative size threshold for division is also operative in other species.  For exponential growth, this feature corresponds to the growth rate and the division time varying linearly with each other as the temperature changes; the ratio of the joint size of the daughter cells to the mother cell sets the proportionality constant.  Regardless of the threshold mechanism, the division time distribution should scale with the mean so long as only a single timescale contributes to the stochastic dynamics \cite{iyer-biswas-fk}.  Clearly this is not always the case:  the DNA replication time is distinct from the doubling time for {\it E.\ coli} under favorable nutrient conditions, as recently modeled \cite{Ariel PRL}. However, a single timescale could still characterize growth and division of {\it E.\ coli} in minimal media, when replication and division frequency are approximately equal.

The fact that effective activation energies for population growth are generally in the range of individual enzyme-catalyzed reactions \cite{1943-sizer} suggests that the SHC applies broadly.  Stochastic exponential growth implies growth dynamics with a single timescale. The division time distribution scales with its mean when the exponential timescale is proportional to the mean division timescale \cite{iyer-biswas-fk}.   Clearly this need not always be the case:  the DNA replication time is distinct from the doubling time for {\it E.\ coli} under favorable nutrient conditions, as recently modeled \cite{2014-amir-vn}; it remains to be determined if these two timescales change proportionally when temperature is varied. However, a single timescale could still characterize growth and division of {\it E.\ coli} in minimal medium, when replication and division frequency are approximately equal.  The relative size threshold for division is a new paradigm for how cell size can inform cell division.   For exponential growth, this feature is related to the growth rate and the division time varying linearly with each other as the temperature changes; the ratio of the joint size of the daughter cells to the mother cell sets the proportionality constant.  

\subsection{Molecular basis}

%The factors that determine the absolute value of the single-cell exponential growth rate also remain to be determined. The scaling behaviors reported here are insensitive to the absolute value of $\k$ and its variation with external parameters (as in Fig.\ \ref{fig-ig}A). 

{How precisely molecular interactions set the growth rate, how they couple to the divisome and cell wall synthesis machinery, and how the associated network gives rise to SHC dynamics} remain to be determined for each exponential growth condition.  We caution against interpreting exponential growth of cell size as necessitating a spatially uniform distribution of active growth sites on the cell since polar growth of single {\it A.\ tumefaciens} cells has been observed to be super-linear and is potentially exponential~\cite{2012-brown-la}. In [48], we show that complex autocatalytic networks can be systematically reduced to effective SHC models. Therefore,  the scaling laws discussed here would persist even when the biochemical networks that govern growth depend on condition. Previous studies argued for an $N=2$ cycle composed of the global production of metabolic proteins at a rate proportional to the numbers of ribosomal RNA and vice versa \cite{1958-schaechter-ly, 1979-herendeen-qf}, leading to a constant ratio of the two species \cite{2010-hagen-fk}. However one should not take this model literally because metabolic proteins do not directly produce ribosomes. Judicious use of antibiotics and alternative growth media, as in \cite{2010-scott-vn, 2011-scott-kx}, together with our single-cell technology could provide important clues into contributing biochemical reactions for a given condition.
%\cite{iyer-biswas-fk} 
%We note that the mechanics of cell wall synthesis must be coupled to cell growth via the regulation of number density of active growth sites by a component of the SHC \sib{(REF: Ariel)}. 

%That said, our measurements cannot resolve whether the cell senses its size or the time since the last division. However, we note that the coefficient of variation for the division time distribution is $\approx$13\%, compared with the previously mentioned $\approx$8\% for the relative area distribution.

\section{Conclusions}

The preponderance of recent work on bacterial growth in single cell studies has focused on bottom-up explorations of specific regulatory networks \cite{2010-curtis}, and some simple empirical laws connecting global gene expression patterns with the growth state of the cell have emerged~\cite{2011-scott-kx, 2012-velenich-vn}. The complementary, top-down approach of utilizing observations at the organismic level to deduce constraints on microscopic models~\cite{1962-schaechter-dz, 1962-koch-kl, 1958-schaechter-ly, 1968-donachie-mi} has been less popular in the last few decades. In this paper we have taken the latter approach but now with the advantage of being able to acquire and analyze large datasets.  We have observed robust scaling laws for cell growth and division, in addition to the observation of exponential growth of mean single cell sizes. 
To summarize, these single-cell scaling laws are as follows.
\begin{enumerate}
\item{The growth law is exponential during balanced growth under favorable nutrient conditions.} 
\item{The mean division time is proportional to the inverse of the mean growth rate.}
\item{The size of the cell at division is proportional to the initial size of the cell.}
\item{The mean-rescaled division time distribution is temperature invariant.} 
\item{The mean-rescaled cell size distribution from all times and temperatures is invariant.} 
\item{The coefficient of variation of cell sizes, for a given initial cell size, scales as the square-root of time.}
\end{enumerate}
To the best of our knowledge, the SHC is the simplest model that captures all these behaviors, not just the trends of the means but also those of the fluctuations. {We additionally showed that the averaging of the rate constants in the Hinshelwood cycle can account for  the energy scale implied by the temperature dependence of the mean growth rate, which is on the order of a single enzyme-catalzyed reaction's activation energy. However, we emphasize that the variation in Fig.\ \ref{fig-ig}A is not itself a scaling law and only serves to ``calibrate'' how the absolute unit of time (the mean division time) varies with the external parameter (temperature); the scaling laws enumerated above are insensitive to the form of the temperature dependence.}  Our data and the SHC \cite{iyer-biswas-fk} show that stochastic growth and division are governed by a single timescale, which, in turn, depends on the growth conditions.  This simple design principle is  unexpected given the complexity of a whole organism.

\section{Methods}

%\subsection{Summary of experimental procedure} 
The experimental techniques developed here enable studies of individual non-interacting cells in well-controlled environments for $>$100 of generations. We have generated a strain in which the only functional copy of the {\em holdfast synthesis A (hfsA)} gene, which controls features required to adhere to surfaces, is integrated at a single chromosomal locus under the control of an inducible promoter. As a result, we can initially induce holdfast production until we have the desired numbers of cells sticking to the glass surface of the microfluidic device and then flow away the remaining cells. Once the experiment commences, the inducer is removed and newborn daughter cells, upon differentiation, do not express functional {\em hfsA} and are thus unable to stick and are flowed away. This prevents the crowding of the fields of view that occurs in typical experiments with exponential growth. 

We use phase-contrast imaging  to accurately measure growth frame-by-frame, instead of just division events. In a typical experiment, data from 20 unique fields of view are acquired at a rate of each field per minute (roughly 100,000 images in three days). Finally, we have developed custom  software using a combination of Matlab and Python for automated image processing, which is necessary for extracting quantitative information from these extensive data ($\sim$10$^6$ images for each temperature studied).  See SI Text Section 1 for further details.

\section{Acknowledgments}

We thank Aretha Fiebig,  Ariel Amir, Rutger Hermsen, Gurol Suel, Kingshuk Ghosh, Matt Scott, Terry Hwa, William Loomis and Leo Kadanoff for insightful discussions.  We thank the  National Science Foundation (NSF PHY-1305542) and the W.\ M.\ Keck Foundation for financial support.  We also acknowledge partial financial and central facilities assistance of the University of Chicago Materials Research Science and Engineering Center, supported by the National Science Foundation (NSF DMR-MRSEC 0820054).  

\section{Author Contributions}

S.I.B., Y.L.,  A.R.D. and N.F.S. conceived of and designed the experiments. S.I.B. and C.R.W. performed the experiments. C.R.W. designed custom software for image analyses. S.I.B., Y.L. and J.T.H. did the cloning. S.I.B.,  C.R.W. and K.L. performed data analyses. S.I.B. observed scaling behaviors reported, constructed the theoretical model and performed calculations. S.I.B., G.E.C. and A.R.D. tested the model. S.C. oversaw the molecular biology and contributed reagents, materials and analysis tools for cloning and cell culturing. S.I.B., A.R.D. and N.F.S. wrote the paper. All authors discussed the results and commented on the manuscript.

%\pagebreak
%\end{article}
%{\bf Exponential growth at different temperatures:}
%%%%%%%%%%%%%%%%%%%%%%%%begin%%%%%%%%%%%%%%%%%%%%%%%%

%\bibliographystyle{pnas-bolker}
%\bibliography{iyerbiswas2013a,added,temp}

\newpage
%%%%%%%%%%%%%%%%%%%%%%%%%end%%%%%%%%%%%%%%%%%%%%%%%%

\clearpage

%{\bf Size-ratio thresholding for division: }
%%%%%%%%%%%%%%%%%%%%%%%%begin%%%%%%%%%%%%%%%%%%%%%%%%
%%%%%%%%%%%%%%%%%%%%%%%%%end%%%%%%%%%%%%%%%%%%%%%%%%
\clearpage
%%%%%%%%%%%%%%%%%%%%%%%%begin%%%%%%%%%%%%%%%%%%%%%%%%
%%%%%%%%%%%%%%%%%%%%%%%%begin%%%%%%%%%%%%%%%%%%%%%%%%
\clearpage

%%%%%%%%%%%%%%%%%%%%%%%%begin%%%%%%%%%%%%%%%%%%%%%%%%
%%%%%%%%%%%%%%%%%%%%%%%%begin%%%%%%%%%%%%%%%%%%%%%%%%
\clearpage

%%%%%%%%%%%%%%%%%%%%%%%%begin%%%%%%%%%%%%%%%%%%%%%%%%
%%%%%%%%%%%%%%%%%%%%%%%%begin%%%%%%%%%%%%%%%%%%%%%%%%
\clearpage

\onecolumngrid

%\vspace{1cm}
\begin{center}
{\bf\Large Supplemental Information}
\end{center}
%\vspace{0.1cm}
\setcounter{secnumdepth}{3}  
\setcounter{section}{0}
\setcounter{equation}{0}
\setcounter{figure}{0}
\renewcommand{\theequation}{S-\arabic{equation}}
\renewcommand{\thefigure}{S\arabic{figure}}
\renewcommand\figurename{Supplementary Figure}
\renewcommand\tablename{Supplementary Table}
\newcommand\Scite[1]{[S\citealp{#1}]}
\makeatletter \renewcommand\@biblabel[1]{[S#1]} \makeatother

\section{Experimental Methods}
%\section*{Details of the experimental setup and image processing}
%\Should I have separate subsections for cloning, experimental set up and image analysis.

\subsection{Cloning of the mutant strain, FC1428} {\em C. crescentus} strain CB15 naturally adheres to surfaces via an adhesive
polysaccharide termed a holdfast; production of holdfast requires the {\em hfsA} gene~\cite{2003-smith-uq-s}. Strain NA1000 is a laboratory-adapted relative
of CB15 and bears a frameshift mutation in {\em hfsA}, rendering cells non-adhesive~\cite{2010-marks-vn-s}. We cloned the functional {\em hfsA(CB15)} allele into
pMT-$862$~\cite{2007-thanbichler-zr-s} and integrated it into the
NA1000 chromosome at the {\em vanA} locus, under a vanillate-inducible promoter. The
resultant strain, FC1428, only gains the ability to adhere to surfaces when exposed to vanillate. Cells are induced with $0.5 $ mM vanillate for
$3$ h before introduction into the microfluidic device; they are then allowed to
adhere to the glass interior of the device. Vanillate-free media is then flowed over
the cells for the remainder of the experiment; induction of {\em hfsA(CB15)} does not
occur in newborn cells, which do not adhere and are thus washed out of the microfluidic chamber.
This Òinducibly-stickyÓ strain allows for long experimental run times, as a constantly
adherent strain would rapidly crowd the field of view with daughter cells produced over many generations.

\subsection{Growth protocol} For each experiment, individual colonies of FC1428 were selected from a fresh PYE-agar plate containing kanamycin ($5$ $\m$g/ml) and grown overnight in PYE medium in a $30^{\circ}$C roller incubator, taking care to ensure that the culture was in log phase. This culture was diluted to $\mbox{OD}_{660} =0.1$ with fresh PYE and $0.5$ mM vanillate and was induced for $3$ h prior to being loaded onto the microfluidic channel in the previously temperature-stabilized  chamber. PYE, Peptone Yeast Extract, is a complex medium and its detailed composition is provided in~\cite{1991-ely-ys-s}.

%Two computer-controlled syringe pumps (PHD2000, Harvard Apparatus), also in the temperature controlled chamber, were used to pump thermally equilibriated PYE media at a constant flow rate of $7\,\, \m$L/min.

\subsection{Microfluidic device and single-cell assay}
\begin{figure}[h]
\begin{center}
\resizebox{10cm}{!}{\includegraphics[trim= 0cm 8cm 0cm 4cm, clip = true]{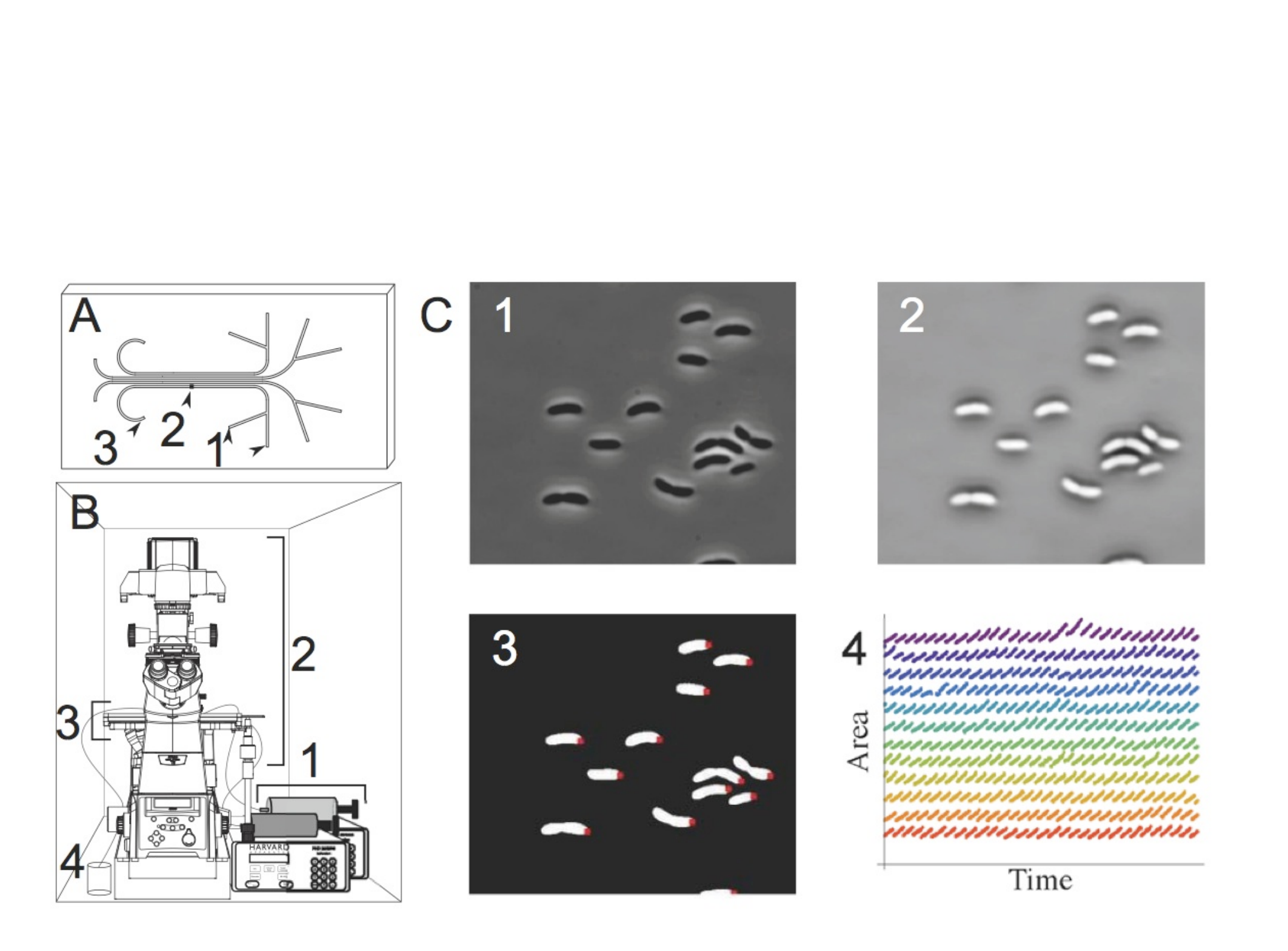}}%[trim=tcm lcm bcm rcm, clip=true, angle=-90]
\caption[Schematic of the experimental setup]{{\bf Schematic of the experimental setup.} (A) The microfluidic device (A1) creates a constant perfusion environment within the channel where imaging occurs (A2); there is continuous fluid exchange through the output (A3). It consists of four individual channels, which are connected to capillary tubing to create a sealed environment. Inputs of two different media may be connected at the upstream end. (B) The experimental apparatus. Each  syringe is attached to a separate syringe pump to allow mid-experiment switching between media (B1); images are obtained using a Nikon Ti-E microscope with auto-focus (B2), which compensates for focal drift as the robotic XY stage holding the microfluidic device (B3) moves between multiple fields of view within the microfluidic channel during the course of long-term experiments. (B4). Each component is controlled by a custom LabVIEW program that completely automates the process of data acquisition after the experiment has been setup.  (C) Image processing workflow. An example of the raw data (1024 pixel $\times$ 1024 pixel), a phase contrast image is shown in (C1). Each  image is then processed with the goal of accurately and robustly detecting cell edges (C2). Features are then identified (C3): the processed images are thresholded to extract cell areas (white), and the point on each cell perimeter closest to the holdfast (red) is assumed to represent a near-stationary point and used to track cells (i.e., to maintain cell identity between every frame of the movie). A typical cell trajectory obtained with the above algorithm is shown (C4).}
\label{fig-schematic}
\end{center}
\end{figure}

See Figure~\ref{fig-schematic} for details of the microfluidics, optics, and image processing aspects of the experimental setup. Y-shaped microfluidic channels were fabricated and prepared as described in~\cite{2012-lin-uq-s}. After thermal equilibration, the FC1428 bacterial cell culture was loaded into a single channel and incubated for 1 h. Typically enough cells stuck to the glass surface of the device after a 1 h period of incubation for the subsequent imaging experiment. The remaining cells (i.e., those that were not adherent) were then washed off in the laminar flow of the microfluidic device. Two computer-controlled syringe pumps (PHD2000, Harvard Apparatus) pumped thermally-equilibrated PYE media through the channel at a constant flow rate ($7\,\,\m$L/min).

\subsection{Time-lapse microscopy}  The imaging process was automated such that the imaging,  stage positioning, illumination, syringe pumps and readout from the array detector were fully computer controlled and could operate autonomously throughout experiments of many days. Time-lapse single-cell measurements were performed on an  inverted microscope (Nikon Ti Eclipse) equipped with a motorized sample stage and a controller (Prior Scientific ProScan III). Phase-contrast microscopy was performed with a Nikon Plan Fluor 100X oil objective and a mercury fiber illuminator (Nikon C-HGFI). A computer-controlled shutter (Lambda SC) was used to coordinate light exposure and image acquisition. The image was collected on an electron multiplying charge coupled device detector (EMCCD, Andor iXon+ DU$888$ 1k x 1k pixels). To ensure thermal stability, the microscope and syringe pumps were enclosed by a homemade acrylic microscope enclosure $(39" \times 28" \times 27")$ heated with a closed-loop regulated heater fan (HGL419, Omega). A uniform temperature was maintained by a proportional integral derivative temperature controller (CSC32J, Omega) coupled with active air flow from two small-profile heater fans inside the enclosure. For experiments carried out below  $20^\circ$C, the  temperature in the entire room was lowered to $6^\circ$C and the aforementioned enclosure that includes the microscope was heated. Phase-contrast images of multiple fields-of-view were recorded at 1 frame/min and the focus adjusted automatically using the built-in ``perfect focus system'' (Nikon PFS). A Virtual Instrument routine (LabView 8.6, National Instrument) was used to control all components (sample stage, autofocus, pumps, EMCCD, and shutter) and to run the experiment for extended periods of time (5-12 days). 

\subsection{Image analysis and construction of growth curves} The acquired phase-contrast images were processed by identifying each {\em C. crescentus} cell in Matlab (MathWorks) and tracking the cells over time using custom code written in Python. The cross-sectional areas of each cell measured through a sequence of images were used to determine growth curves. From these data division events were identified. We chose to only include  cells that divided for more that $10$ generations in the analysis, as a selection criterion for further analysis.

%It is important to note that the datasets only included the original  set of stalked cells so that the population was generationally homogeneous.

\section{Cell size determination and precision}

Typical Gram-negative bacteria have cylindrical rotational symmetry around their anterior-posterior axes.  In {\em C. crescentus}, the symmetry is around a curved axis since the cells are crescent shaped.  As shown in Fig.\  S3, we have verified that the growth of the cell is predominantly along the longitudinal direction, by evaluating the curved mid-cell axis (the bisector of the observed area of the cell); this length itself grows with the same exponential growth rate as we deduce from the area. However, quantifying cell size by the straight-line length joining the anterior-posterior extremities of the cell instead would lead to an accumulation of errors because it ignores the inhomogeneous width of the cell perpendicular to this line, in the plane of observation.  We use area because it obviates this problem and affords us an order of magnitude better precision. We expect the area to reflect the volume faithfully because cells are cylindrically symmetric and their lengths grow exponentially with the same time constant as the area.

Using a combination of thresholding the absolute intensity and ridge detection algorithms, the (pixellated) boundary of each cell was identified, frame by frame.  Cell area was quantified by counting the total number of pixels inside the boundary for each cell in each frame. We compute the precision of our measurements in several different ways. First, we vary the threshold for cell edge detection over a $10\%$ range ($\pm 5\%$ of the value used for all image analysis) and find that that the area of each cell is changed by $\approx$2\% at the beginning of each cell cycle; this number decreases further as the cell grows. Second, we perform control experiments with more frequent sampling (30 frames/min) and use bootstrapping methods to estimate the error bars on the precision of our single-cell measurements in our experiments, which are performed at 1 frame/min. Third, we examine the fluctuation in the areas of cells that do not grow during the course of the experiment but are not dead (a condition that is controlled by the media), at 1 frame/min.  The measurement uncertainty of a single-cell area is $<$2\%.  Since we obtain between 4,000 and 16,000 growth curves at each temperature, the ensemble averaged mean area at a given instant of time has an uncertainty of $<$0.03\%.

The division times are taken to be the minima of the area vs.\ time curves. We estimate that the error in division times is less than 2 min (twice the inverse frame rate). Since the Coefficient of Variation (ratio of the standard deviation to the mean) of the division time distributions at all temperatures is $\approx$0.13 (see following section), the standard error in the mean division times at each temperatures (with 4,000-16,000 points) is 0.01-0.03 min.

\section{Determination that the growth law is exponential}

\subsection{Fitting individual trajectories}
The Langevin model for stochastic exponential cell size growth is given by {Eq.\ 4} of the main text and is used to find the correct procedure for ensemble averaging the growth curves to obtain the time evolution of the  mean cell size, i.e., the growth law(s). Upon integrating this equation,
%%%%%%%%%%%%%%%%%%%%%%begin%%%%%%%%%%%%%%%%%%%%%%
\begin{align}%\label{eq-}
e^{-\k(T)t/2} \sqrt{a(t; T)} - \sqrt{a(0;T)} = \frac{1}{2} \int_{0}^{t}d t' e^{-\k(T)t'/2} \, \eta(t').
\end{align}
%%%%%%%%%%%%%%%%%%%%%%%end%%%%%%%%%%%%%%%%%%%%%%
Thus, the time evolution of the square of the ensemble averaged mean of the square root of the size is exponential:
%%%%%%%%%%%%%%%%%%%%%%begin%%%%%%%%%%%%%%%%%%%%%%
\begin{align}%\label{eq-}
\le\la \sqrt{a(t; T)}\ri\ra^{2} = \le\la \sqrt{a(0;T)}\ri\ra^{2}e^{\k(T)t}.
\end{align}
%%%%%%%%%%%%%%%%%%%%%%%end%%%%%%%%%%%%%%%%%%%%%%
Using this result, at each temperature, we fit the growth data for each generation,  $\sqrt{a(t;T)}$ vs. $t$, with the best exponential fit, to find $\k/2$ and thus $\k$.

%\subsection{Ergodicity and intergenerational correlations}
Since the mean and standard deviation of the growth rates and division times evaluated by considering different generations of the same cell were  equal to the same quantities evaluated across different cells at a given generation, the ergodic condition that ensemble averaging  equals generational averaging holds for these data. Therefore we do not see a systematic change in the reproductive output of a given cell from generation to generation, under these growth conditions. 

A related issue is that of intergenerational correlations in these quantities. We find that there is a small but observable anti-correlation between the initial size of the cell and its division time at the end of that generation (but no correlation between the initial size and the growth rate) at all temperatures. This mild anti-correlation serves to restore the (absolute) size of a cell to the ensemble average and prevents ``runaway'' cells, i.e., larger (smaller) than average cells from getting progressively larger (smaller), compared to the ensemble mean, due to noisy relative size thresholding at division.

\subsection{Distinguishing between functional forms}\label{discrim}
What functional form best fits the ensemble averaged mean growth law, i.e., the increase of mean cell size with time in balanced growth conditions, has been debated; the two main  contenders are the linear and exponential forms~\cite{2006-cooper-rm-s, 1991-cooper-sf-s, 1968-kubitschek-lq-s,2001-koch-yg-s}. 

An important reason why  ascertaining the growth law, beyond reasonable  doubt, has been an experimental challenge is because extraordinary (statistical) precision  is required to distinguish an exponential from a straight line when each growth period is  less than the time constant of the exponential. This can be seen by estimating the minimum precision required for discriminating between the two functions, by considering the geometrical aspects of exponential and linear curves for a given mean growth period, $\la \t \ra$, and a relative division threshold $\th \equiv \la a(\t) \ra/\la a(0)\ra$~\cite{2009-tzur-uq-s} {(see  Fig.\ 2B (main text) and Fig.~\ref{fig-thresh}).} The time at which the exponential curve deviates most from the straight line is then found to be
%%%%%%%%%%%%%%%%%%%%%%begin%%%%%%%%%%%%%%%%%%%%%%
\begin{align}%\label{eq-}
\t_{m} = \la \t \ra \frac{1}{\ln{\th}}\,\ln\le[\frac{\th-1}{\ln \th} \ri].
\end{align}
%%%%%%%%%%%%%%%%%%%%%%%end%%%%%%%%%%%%%%%%%%%%%%
The maximal difference between the predicted sizes for the exponential and linear models ($ \D a_{max}$) is thus the difference between the sizes predicted using each model  at time $\t_{m}$:
%%%%%%%%%%%%%%%%%%%%%%begin%%%%%%%%%%%%%%%%%%%%%%
\begin{align}%\label{eq-}
\D a_{max} = \la a(0) \ra \le[ 1 + (\th-1)\le\{\frac{\t_{m}}{\la \t \ra} -\frac{1}{\ln \th} \ri\}\ri].
\end{align}
%%%%%%%%%%%%%%%%%%%%%%%end%%%%%%%%%%%%%%%%%%%%%%
Thus the minimum precision of measurement required to distinguish between these models is determined by whether $\D a_{max}  \gg \s(\t_{m})$ or not, where $\s(t)$ is the standard deviation in $a$ observed at time $t$. Scaling $\D a_{max}$ by the predicted size at $\t_{m}$ for the exponential model and defining $ f(\th) \equiv {(\th-1)}/{\ln \th}$, we thus arrive at the minimum precision required for distinguishing between the two models.
%%%%%%%%%%%%%%%%%%%%%%begin%%%%%%%%%%%%%%%%%%%%%%
\begin{align}%\label{eq-}
\parbox{0.75in}{\textrm{minimum precision}}=\frac{\la a(0) \ra {\la \t \ra} \le[ 1 + f(\th)\le( \ln\le( f(\th)\ri) -1 \ri)\ri]}{\la a(0) \ra \le[ {\la \t \ra} + {\t_{m}}(\th -1)\ri] } =  \frac{1 + f(\th) \le[ \ln\le(f(\th) \ri) -1 \ri]}{  1 + f(\th)\,\ln\le( f(\th) \ri)}.
\end{align}
%%%%%%%%%%%%%%%%%%%%%%%end%%%%%%%%%%%%%%%%%%%%%%
The required precision  is $\approx$4$\%$ for a division size ratio of $\th = 1.8$,  as is observed in our experiments. Since error in our mean area measurements is less than $0.03 \%$, we can indeed unequivocally distinguish between exponential and linear growth. 

To quantify the goodness of fit for both the exponential and linear fits, and to establish that the statistically preferred model is the exponential one, we use the following prescription. We recall that the ensemble averaging procedure that correctly accounts for the cancellation of the noise contribution from $\eta(t)$, for the model of stochastic growth proposed, is to find the root-mean-square of the area, $\le\la \sqrt{a(t)}\ri\ra^{2}$ at each observation time (a noise model with additive noise or linear multiplicative noise is contradicted by the scaling of cell size distributions observed). In this ensemble averaging procedure, no ad hoc subtraction of or division by the initial size to de-trend the noise is necessary. If the growth law were linear rather than exponential, then $\le\la \sqrt{a(t)}\ri\ra^{2}$ should fit better to a model that is of the form $c\, t + d$, where $c$ and $d$ are parameters of the linear model. 
The exponential  fit has $\chi^{2}\approx 50$ for all temperatures (Table~\ref{table1}), compared with  $\chi^{2}\approx1000$ for the best linear fits (Table~\ref{table1}). Since both models, exponential and linear, have the same number of degrees of freedom, two fitting parameters each (i.e.,  the mean initial size and the mean growth rate), the Akaike Information-theoretic criterion index (AIC)~\cite{2002-burnham-fk-s}  for each is simply given by its respective $\chi^{2}$ value. Clearly the $\chi^{2}$ value for the exponential model is much smaller than that for the linear growth model. However, we can use the AIC      to determine the relative likelihood that the linear model  is the correct description of data, not the exponential. Using $\exp{\le[ (\mbox{AIC}_{\rm exp} - \mbox{AIC}_{\rm lin})/2\ri]}$, we find that it  ranges from  $10^{-70}$ to $10^{-500}$ for the different temperatures (the variability in the value coming from the differences is total numbers of growth curves at each temperature).  Therefore statistical measures of model selection overwhelmingly favor the exponential form. 

\begin{figure}[ht]
\begin{center}
\resizebox{16cm}{!}{\includegraphics[trim=0cm 6cm 0cm 4cm, clip=true]{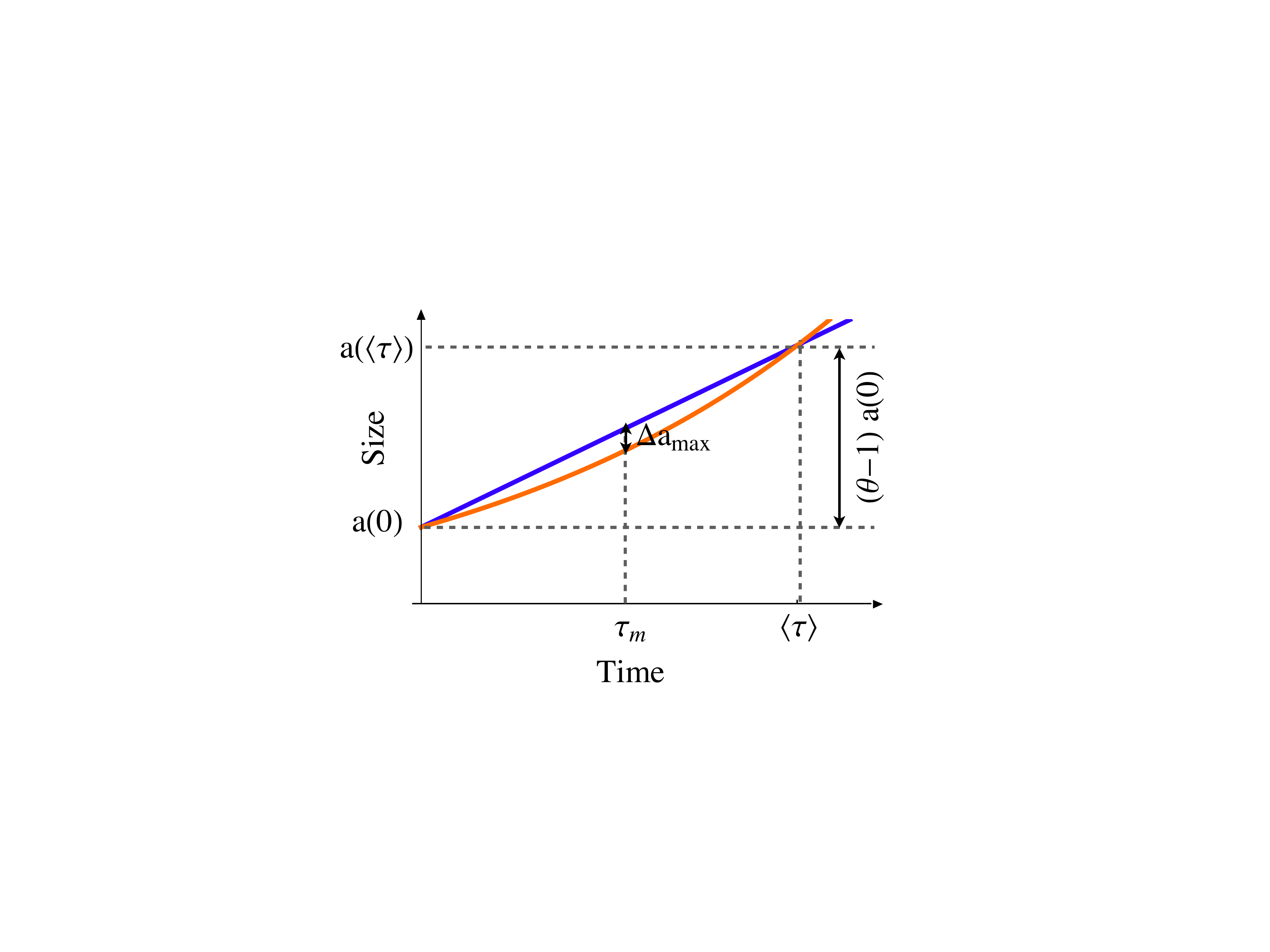}}
\caption[Schematic illustrating the challenge of discriminating exponential and linear models.]{{\bf Schematic illustrating the challenge of discriminating exponential and linear models.}
For a cell growing from an initial size, $a(0)$ to a multiple $\th$ of the initial size, i.e., $\th a(0)$, the linear (blue) and the exponential  (red) fits  (both passing through the initial and final points)  maximally differ at a time $\t_{m}$ and the magnitude of the maximal difference is $\Delta a_{max}$. The measurement precision has to be better than $\Delta a_{max}$ for model selection (between linear and exponential) to be feasible.  See SI Text Section \ref{discrim} for discussion.
}
\label{fig-compare}
\end{center}
\end{figure}

We note that the residuals for the exponential fit in Fig.~\ref{fig-compare}C have additional structure, not fully explained by a model that assumes a constant (time-independent) mean growth rate. We believe that the systematics in the residuals for the exponential fit suggest that there may be a small growth phase (cell age) dependence to the growth rate, reflecting specific underlying growth/division processes, such as restructuring of the cell for formation of end caps and the constricting of the division plane; this is an interesting avenue for future enquiry. Here, we  have used the constant growth rate model because it is the most economical model to explain the overwhelming majority of observations.
We thus conclude that the growth law for these cells,  under the conditions described in the text, is exponential. 

%It is worth noting that as seen in \sib {Fig. 5B} (main text), the diffusive contribution in the Langevin equation [Eq. 4 of main text] is nearly two orders of magnitude smaller than the exponential drift. So the noise model, i.e, whether the multiplicative noise is square-rooted or not (and hence the particulars of which ensemble averaging scheme is the self-consistent one) turns out to be inconsequential for determining whether the exponential fit is better or the linear one. The exponential fit is always found to be overwhelmingly better, for a given ensemble averaging scheme.

We note that a formal comparison with other growth laws is also possible.  The exponential fit  compares  favorably with a quadratic function (the simplest higher-order polynomial) too. Geometric considerations similar to those detailed above indicate that the minimum precision required to discriminate between exponential and quadratic growth laws is $\approx$0.1$\%$, which is within our statistical precision. The best fits for the quadratic have coefficients for the quadratic term that are approximately equal (within a factor of 1.2-1.5 times) to the quadratic coefficient of the series expansion of the exponential function. To quantify the statistical significance of the goodness of each fit, we have used the Bayesian Information Criterion (BIC)~\cite{1978-schwarz-ve-s}, a common information-theoretic measure for weighing models with different numbers of fitting parameters (the quadratic has one additional free parameter over the exponential). The BIC for the quadratic fit is greater than  that for the exponential by more than 11, which is very strong evidence against the quadratic.  In summary, the quadratic fit is comparable in quality to the exponential fit but has an additional free parameter, and we thus favor the exponential. 

%A geometrical estimate, similar to one detailed above, for the comparison between exponential and linear growth laws, indicates that the minimum precision required to discriminate between these two growth laws is $\approx$0.1$\%$. Since our statistical precision for the observed mean growth is $\approx$0.03$\%$, we can make this comparison. The best fits for the quadratic have coefficients for the quadratic term which are approximately equal (within a factor of $1.2$--$1.5$ times) of the expected quadratic coefficient, using a (polynomial) series expansion of the exponential function. This result provides further evidence that the exponential growth law is the more parsimonious model to reasonably fit the data. To quantify statistical significance of the goodness of each fit, we have used the Bayesian Information Index (BIC)~\cite{1978-schwarz-ve-s}, which is the preferred IC for discriminating between models with different numbers of fitting parameters. (The quadratic has one additional free parameter over the exponential.) The BIC for the quadratic fit is greater than the that for the exponential by more than $11$, which is ``very strong evidence against higher BIC''. Using these values, the improbability index that the quadratic model is the correct one (rather than the exponential), is $\approx 10^{-5}$. Moreover, there is no biophysical evidence at present, that we are aware of,  to motivate a multiparameter polynomial growth law in preference to an exponential growth law, under these balanced growth conditions.

\begin{table}[bt]
\centering
\begin{tabular}{| c | c | c | c | c | c | c | c | c |}
%{ |p{.7cm} | p{.7cm} | p{.7cm} | p{2cm} | p{2cm} | p{.7cm} |}%{| c | c | c | c | c | c |ÊÊ}
 \hline 
$\,\,\,T ( ^{\circ}\mbox{C})\,\,\,$ & $\,\,\,\,\,\,\,N_{ens}\,\,\,\,\,\, $ & $\{\la {\sqrt{a_{0}}} \ra^{2} (\,\m m^{2}),$ &ÊÊ$\,\,
\{c \,(\,\m m^{2}),\,\,$&\,\,\,\, $\chi^{2}_{{exp}}$& \,\,\,$\chi^{2}_{{lin}}$&\,\,\,\,{Improbability}\,\,\,\,\\ 
$\,\,\,\,\,\,$ & $\,\,\,\,\,\,\,\,\,\,\, $ & $\la \k \ra \,({min}^{-1}) \}$ &ÊÊ$\,\,
d\, ({min}^{-1}) \}\,\,$&\,\,\,\, & \,\,\,&\,\,\,\, {Index}\,\,\,\,\\ 
\hline \hline
$17$ & $9634$ & $\{1.6, 0.0028 \}$ & $\{ 1.6, 0.005\}$ &ÊÊ$50$&ÊÊ$2200$ &ÊÊ$1.0 \times 10^{-500}$ÊÊÊÊÊÊ\\ \hline
$24$ & $4224$ & $\{1.7, 0.0058\}$ & $\{1.6, 0.012 \}$ &ÊÊ$52$ &ÊÊ$1200$&ÊÊ$1.0 \times10^{-200}$ \\ \hline
$28$ & $4769$ & $\{ 1.6, 0.0075\}$ & $\{ 1.6, 0.015\}$ &ÊÊ$56$ &ÊÊ$1300$&ÊÊ$1.0 \times10^{-300}$ \\ \hline
$31$ & $15240$ & $\{ 1.6, 0.0078\}$ & $\{1.6, 0.015 \}$ &ÊÊ$51$ &ÊÊ$1900$&ÊÊ$1.0 \times10^{-400}$ \\ \hline
$34$ & $13340$ & $\{ 1.6, 0.0099\}$ & $\{ 1.6, 0.019\}$ &ÊÊ$32$ &ÊÊ$1400$&ÊÊ$1.0 \times10^{-300}$ \\ 
\hline
\end{tabular}
\caption[Parameters and goodness of fit measures for exponential and linear models of growth]{{\bf Parameters and goodness of fit measures for exponential and linear models of growth.} Columns are temperature, $T( ^{\circ}\mbox{C})$, the number of growth curves, $N_{ens}$, the exponential model fit parameters $\{\la {\sqrt{a_{0}}} \ra^{2} (\,\m m^{2}) , \la \k \ra \,({min}^{-1}) \}$, the linear model fit parameters, $\{c \,(\,\m m^{2}), d\, ({min}^{-1}) \} $, the $\chi^{2}$ value for the exponential fit, the $\chi^{2}$ value for the linear fit, and the Improbability Index, for the linear fit.}
\label{table1}
\end{table}

\section{Fitting the data}

\subsection{Mean division times}
{The mean values of the division times at 34, 31, 28, 24, and 17$^{\circ}$C are  56, 72, 76, 98, and 201 min, respectively. In the main text we show that, if the individual rates of the Hinshelwood cycle exhibit an Arrhenius temperature dependence (in general, with different activation energies), the overall growth rate (equal to the geometric mean of the individual rates) varies similarly with temperature, with an effective activation energy equal to the arithmetic mean of the individual barrier heights. The argument can be generalized to other functional forms for the temperature dependence of the mean growth rate (or division rate). Specifically, if the individual rates instead follow the Ratkowsky form, $k_{i}(T) \sim (T-T_0)^2$ \cite{1982-ratkowsky-xy-s,1983-ratkowsky-sf-s}, where $T$ is absolute temperature and $T_0$ is a parameter of the empirical relation, we find by calculating the geometric mean of the individual rates that the overall growth rate, $\k(T)$, has the following temperature dependence. 
%%%%%%%%%%%%%%%%%%%%%begin%%%%%%%%%%%%%%%%%%%%
\begin{align}%\label{eq-}
\k(T) &= T^{2}\le[ \le( 1- \frac{\la T_{0} \ra}{T} \ri)^{2} -  \frac{\sigma_{T_{0}}^{2}}{T^{2}} + \mathcal{O}\le( \frac{1}{T^{3}}\ri) \ri].  \\
&\approx (T - \la T_{0} \ra)^{2}.
\end{align}
%%%%%%%%%%%%%%%%%%%%%%end%%%%%%%%%%%%%%%%%%%%
Thus, provided that the standard deviation of the individual values of $T_{0}$ is small compared to their mean value, to leading order, the effective growth rate also scales as a Ratkowsky form, with the effective minimum temperature parameter equal to the arithmetic mean of the individual values, irrespective of the number of steps in the Hinshelwood cycle, up to leading order in temperature.  We note that no restriction on $\Delta E_i$ is required for Eq.\ 3 of the main text to hold in the Arrhenius case.
%In this model, the minimum temperature sets the energy scale in analogy to $\Delta E$ in the Arrhenius model, so the Hinshelwood cycle can generally account for the observed temperature dependence of bacterial growth rates.
}

\subsection{Size distribution}
The distribution of cell sizes, under balanced growth conditions, is predicted to be a gamma distribution \cite{iyer-biswas-fk-s}. We rescale the initial size distributions at all temperatures by their mean values (note that these distributions undergo a scaling collapse and thus have the same shape), and the resulting scaled distributions collapse to a single gamma distribution with a mean of 1. The only parameter of the distribution left to be determined is the (dimensionless) shape parameter; the value that we obtain for it by fitting is 16. Thus we obtain the mean-rescaled initial size distribution, $P(\tilde{a}(0))$, where $\tilde{a}(0) \equiv a(0)/ \la a(0) \ra$.

\subsection{Division time distribution}
\begin{figure}[ht]
\begin{center}
\resizebox{15cm}{!}{\includegraphics[trim=0cm 6cm 0cm 0cm, clip=true]{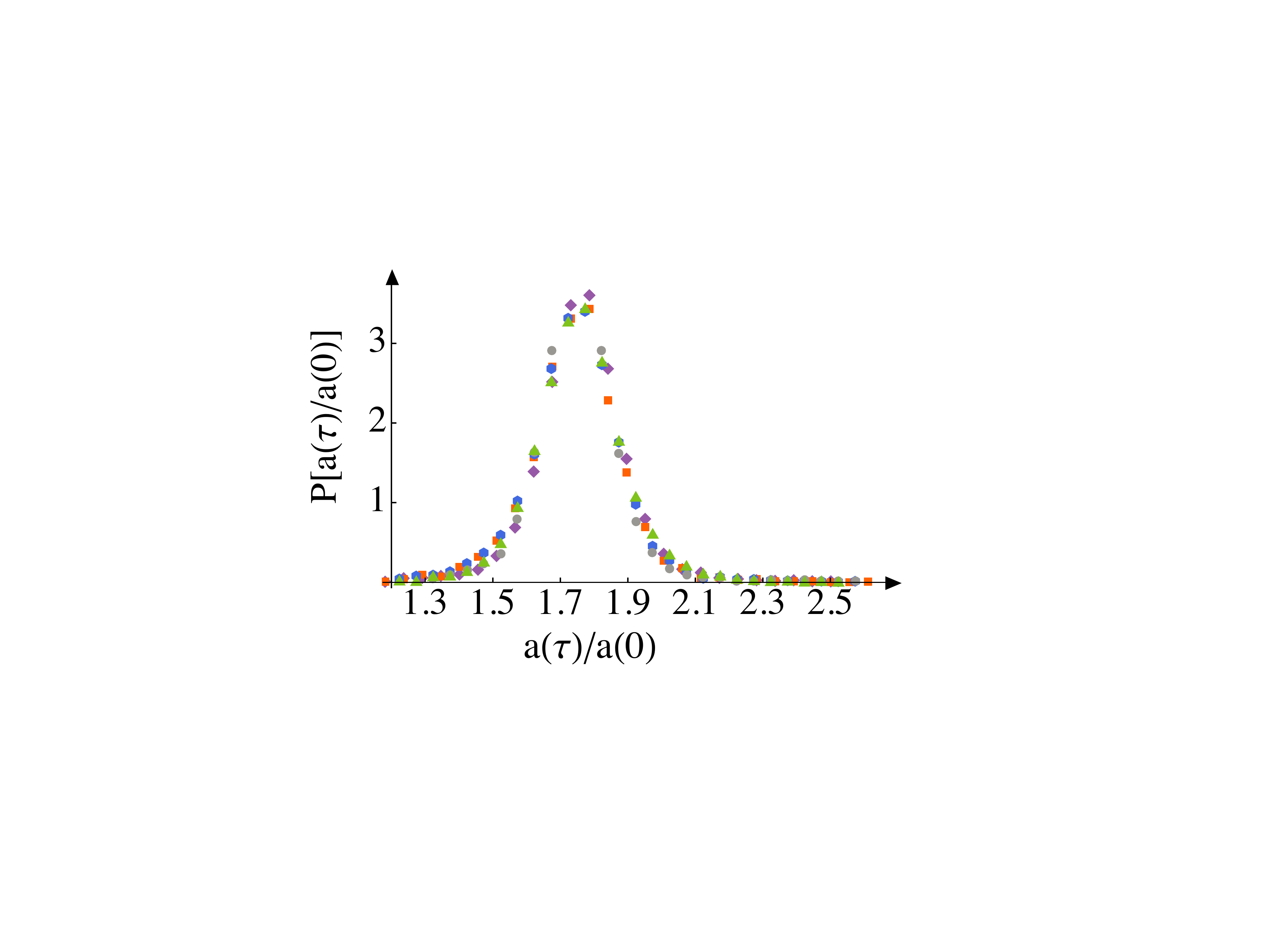}}%[trim=tcm lcm bcm rcm, clip=true, angle=-90]
\caption[Distributions of the relative size threshold at different temperatures]{{\bf Distributions of the relative size threshold at different temperatures.}  The probability distribution of the relative size increase of each cell at division, i.e., the ratio of size at division to initial size, $a(\t)/a(0)$, is shown for all generations and all temperatures in the Arrhenius range (purple, 34$^{\circ}$C; green, 31$^{\circ}$C; orange, 28$^{\circ}$C; blue, 24$^{\circ}$C; gray, $17^{\circ}$C). This plot shows that the distributions undergo a scaling collapse. The mean value is 1.76 and the coefficient of variation is $\approx$8\%.
}
\label{fig-thresh}
\end{center}
\end{figure}

The first passage time distribution (i.e., the division time distribution), for a cell that grows from an initial size, $a(0)$, to when it reaches a multiple $\th$ of its initial size, $\th a(0)$, is a Beta-Exponential distribution \cite{iyer-biswas-fk-s},
%%%%%%%%%%%%%%%%%%%%%%begin%%%%%%%%%%%%%%%%%%%%%%
\begin{align}%\label{eq-}
\mathcal{P}(\t | \tilde{a}(0)) =  \frac{\la \k(T) \ra\, e^{-\tilde{a}(0)\, \la \k(T) \ra \tau}\left(1-e^{-\la \k(T) \ra \,\tau }\right)^{(\theta-1)\tilde{a}(0)}}{{\rm Beta}\le[\tilde{a}(0), \tilde{a}(0)(\theta -1) \ri]},
\end{align}
%%%%%%%%%%%%%%%%%%%%%%%end%%%%%%%%%%%%%%%%%%%%%%
where $\rm Beta$ is the Beta function. Note that $\theta$, the multiple of the initial size to which each cell grows, was observed to be $\approx$1.76, on average (see main text and Fig.~\ref{fig-thresh}). The mean growth rate, $\la \k(T) \ra$, is known from observations at each temperature (Table~\ref{table1}). Moreover, the initial size distribution $P(\tilde{a}(0))$ has also been determined (see above). Therefore, the division time distribution,
%%%%%%%%%%%%%%%%%%%%%%begin%%%%%%%%%%%%%%%%%%%%%%
\begin{align}%\label{eq-}
\mathcal{P}(\t) \equiv \int d \tilde{a}(0) P(\tilde{a}(0)) \mathcal{P}(\t|\tilde{a}(0)),
\end{align}
%%%%%%%%%%%%%%%%%%%%%%%end%%%%%%%%%%%%%%%%%%%%%%
can be computed at each temperature without any additional fitting parameters. 

For the fit in {Fig.\  3B} of the main text, we  restricted ourselves to data $\pm$20\% of the mean growth rate since events outside of this range correspond to biological phenomena not included in the simple model (which assumes a constant growth rate), such as cells that become filamentous. However, these outliers are included in the scatter plots in {Figs.\  2A and B}.  The Coefficient of Variation (ratio of the standard deviation to the mean) of the division time distributions at all temperatures in the Arrhenius range (17-34$^{\circ}$C) is $\approx$13\%.

%\clearpage

\section{Scaling behaviors beyond the Arrhenius range}

\begin{figure}[ht]
\begin{center}
\resizebox{15cm}{!}{\includegraphics[trim=0cm 6cm 0cm 5cm, clip=true]{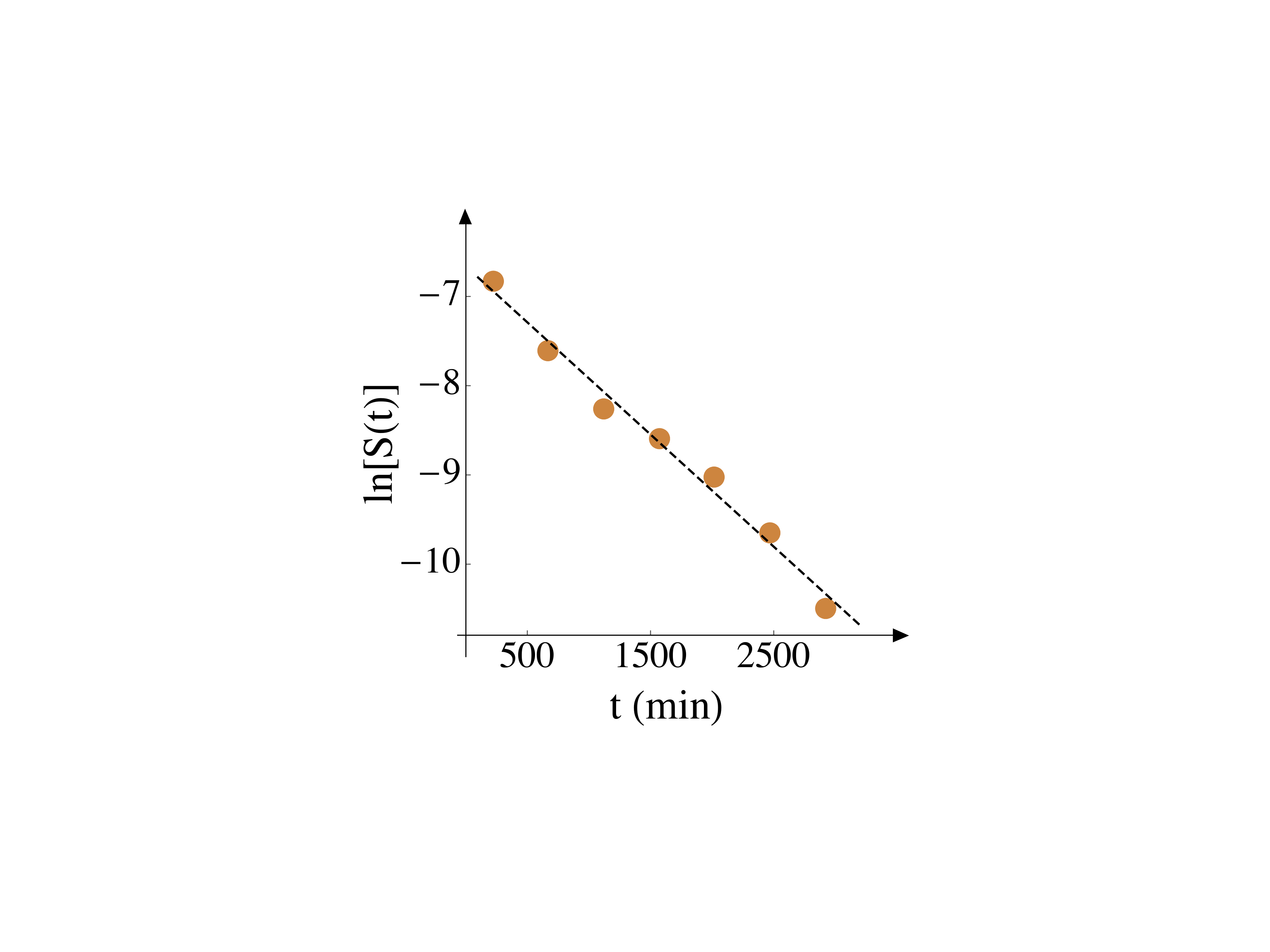}}%inclgphcs[trim=lcm bcm rcm tcm, clip=true, angle=-90]
\caption[The survival distribution at 37$^{\circ}$C]{{\bf The survival probability distribution at 37$^{\circ}$C.}The survival probability, $S(t)$,   is observed to be an exponential distribution (straight line on a log-linear plot); $S(t) \sim e^{-\nu t}$, where $\nu$ is the probability per unit time that a cell dies, fits to $7\%$ per mean duration of the generation of a cell (54 min).  Data are from 241 cells.  ``ln'' denotes the natural logarithm.}
\label{fig-mortality}
\end{center}
\end{figure}

\begin{figure}[ht]
\begin{center}
\resizebox{15cm}{!}{\includegraphics[trim=0cm 6cm 0cm 5cm, clip=true]{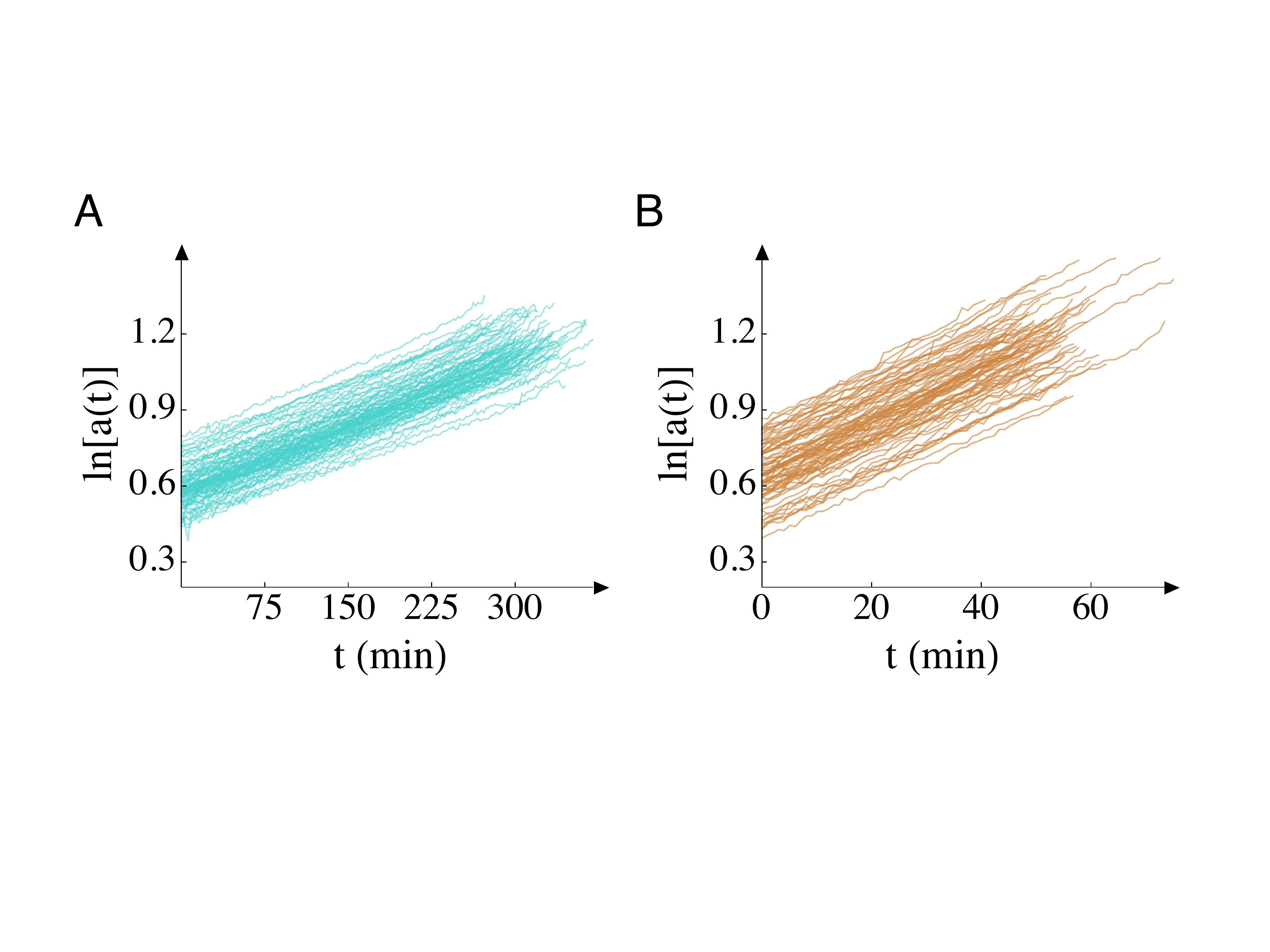}}%inclgphcs[trim=lcm bcm rcm tcm, clip=true, angle=-90]
\caption[Exponential growth at extreme temperatures]{{\bf Exponential growth at extreme temperatures.} The single-cell growth law is observed to be exponential (for surviving cells) even at extreme temperatures: 14$^{\circ}$C (A, cyan) and 37$^{\circ}$C (B, brown). Log-linear plots of the cell sizes as functions of time are shown. Growth data  shown are for 80 generations for each condition. ``ln'' denotes the natural logarithm.}
\label{fig-ext-exp}
\end{center}
\end{figure}

\begin{figure}[ht]
\begin{center}
\resizebox{16cm}{!}{\includegraphics[trim=0cm 0cm 0cm 0cm, clip=true]{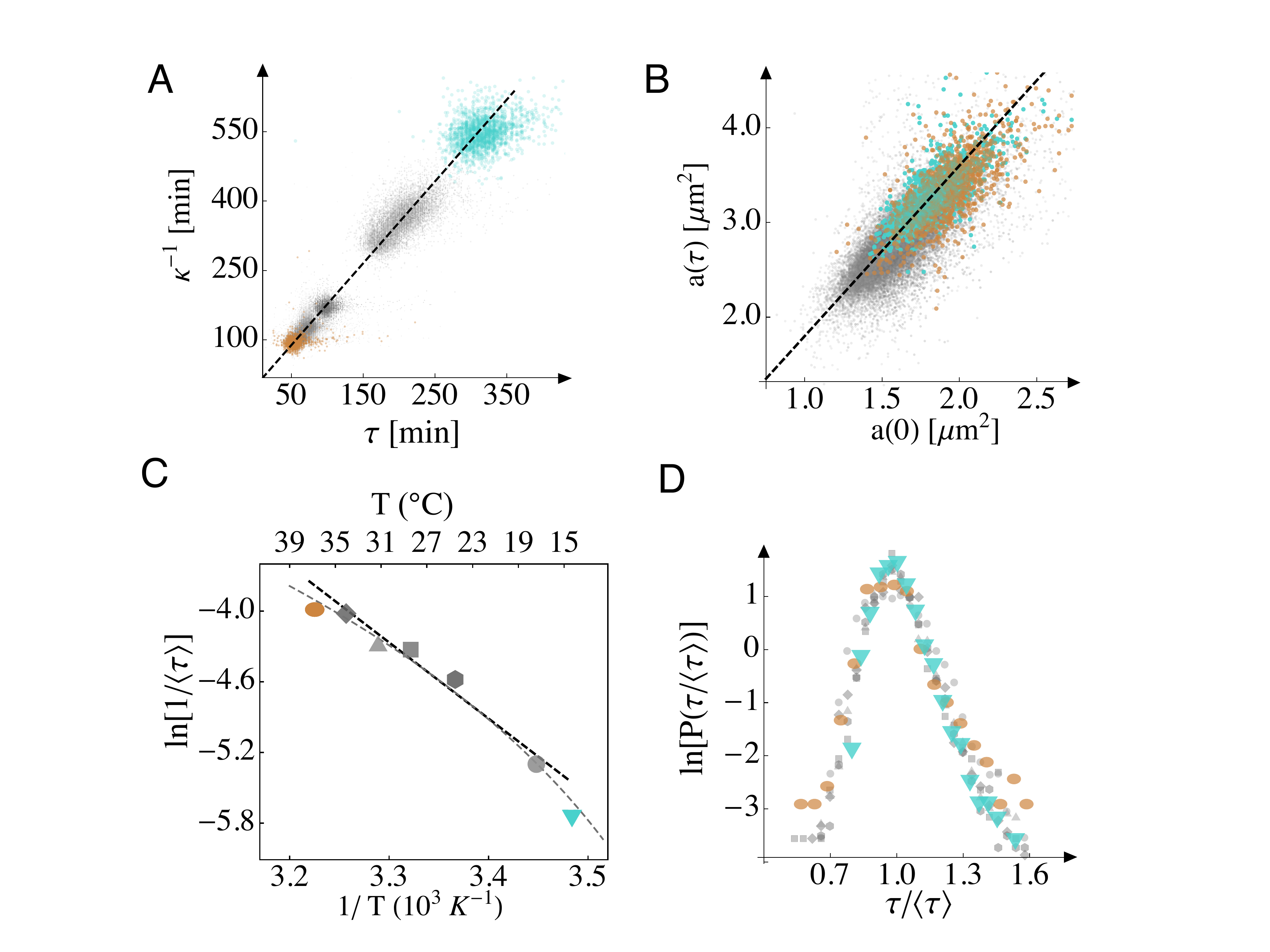}}%inclgphcs[trim=lcm bcm rcm tcm, clip=true, angle=-90]
\caption[ Scaling behaviors at extreme temperatures.]{{\bf Scaling behaviors at extreme temperatures.} Data shown are from $14^{\circ}$C (cyan) and $37^{\circ}$C)(brown), respectively, from 2000 and 4000 growth curves, with 50-200 time points each. Data from temperatures in the Arrhenius range are shown in gray for comparison (compare with Figs. 1, 2 and 3 in the main text). (A) Linear scaling of the division time scale with the growth time scale; the slope of the best fit line (dashed black) is {1.8}. (B) Relative size thresholding of single cells; the slope of the best fit straight line (dashed black) is {1.8}. (C) Breakdown of Arrhenius scaling of the mean division rate, at extreme temperatures. The best fit line for the temperatures in the Arrhenius range (Fig. 3A, main text) is shown for comparison. (D) The mean-rescaled division time distributions at both these temperatures are superimposed on those from temperatures in the Arrhenius range. ``ln'' denotes the natural logarithm.%\sib{REDO}
}
\label{fig-scalings}
\end{center}
\end{figure}

As discussed in the main text, we have performed single-cell experiments at $37^{\circ}$C and $14^{\circ}$C, temperatures that are respectively higher and lower than the Arrhenius range (``normal temperature range''~\cite{1983-ingraham-rm-s}) for the mean growth rate, to investigate scaling behaviors at these extreme physiological temperatures. We have obtained data for between $2000$-$4000$ generations (growth curves) for both conditions. We find that the single-cell growth law remains exponential for both these temperatures (Fig.~\ref{fig-ext-exp}). The mean division time observed at $37^{\circ}$C is $54$ min and at $14^{\circ}$C, $319$ min; in contrast, if they had followed the Arrhenius law ({Fig.~\ref{fig-scalings}C}), these values should have been $\approx$44 min and $\approx$237 min, respectively. Thus, the division rate observed at both temperatures is significantly slower than predicted by the the Arrhenius law. However, the mean growth rate (of surviving cells)  slows down proportionally ({Fig.~\ref{fig-scalings}A}); as a result, $\la \t \ra$ and $\la \k^{-1} \ra$ continue to scale linearly with each other, as they do in the Arrhenius range. Moreover, the initial cell size remains proportional to the size of the cell at division even outside the Arrhenius range; at $37^{\circ}$C the mean value of the relative size threshold is $1.8$ at both temperatures ({Fig.~\ref{fig-scalings}B}).  Further, the mean-rescaled division time distribution from $14^{\circ}$C undergoes the same scaling collapse as the remaining temperatures in the Arrhenius range  ({Fig.~\ref{fig-scalings}D}) but the distribution at $37^{\circ}$C is slightly more noisy with COV $\approx$15$\%$.  We believe that this additional stochasticity, compared to other temperatures, is related  to the onset of cell mortality---we observe significant mortality at $37^{\circ}$C and the increased filamentation rate at this temperature. In Fig.~\ref{fig-mortality} we show that the survival probability, $S(t)$ of a cell at  $37^{\circ}$C is an exponential function of time, $S(t) \sim e^{-\nu t}$. By fitting the observed survival distribution, we estimate that  $\nu$, the probability per unit time that a cell may die, is $7\%$ per mean duration of a generation (54 min). The mean-rescaled cell size distributions from different times, at both temperatures, undergo scaling collapses, as predicted by the SHC. We see an increase in the initial cell size at both extreme temperatures, compared to the temperatures in the Arrhenius range; at present, we do not have an explanation for this observation.

%\bibliographystyle{pnas-bolker}
%\bibliography{supplement-sib2013a.bib}

\clearpage

%%%%%%%%%%%%%%%%%%%%%%%begin%%%%%%%%%%%%%%%%%%%%%%
%\begin{figure}[h]
%\begin{center}
%\resizebox{\textwidth}{!}{\includegraphics[trim=0cm 0cm 0cm 0cm, clip=true, angle=-0]{flochart.jpg}}%inclgphcs[trim=lcm bcm rcm tcm, clip=true, angle=-90]
%\caption[Summary of the flow of the main text]{{\bf Summary of the flow of the main text.} Green boxes denote experimental observations and orange boxes denote experimentally  validated predictions from the model proposed. The  red box encapsulates the conclusions drawn from the observations and the model. \sib{Rearrange logic to reflect the narrative in present main text.}}
%%\label{fig-}
%\end{center}
%\end{figure}
%%%%%%%%%%%%%%%%%%%%%%%%end%%%%%%%%%%%%%%%%%%%%%%
%\pagebreak
%%%%%%%%%%%%%%%%%%%%%%%%begin%%%%%%%%%%%%%%%%%%%%%%%%
%%%%%%%%%%%%%%%%%%%%%%%%%end%%%%%%%%%%%%%%%%%%%%%%%%
\clearpage
%%%%%%%%%%%%%%%%%%%%%%begin%%%%%%%%%%%%%%%%%%%%%%
\begin{figure}[ht]
\begin{center}
\resizebox{12cm}{!}{\includegraphics[trim= 0cm 4cm 0cm 4cm, clip = true]{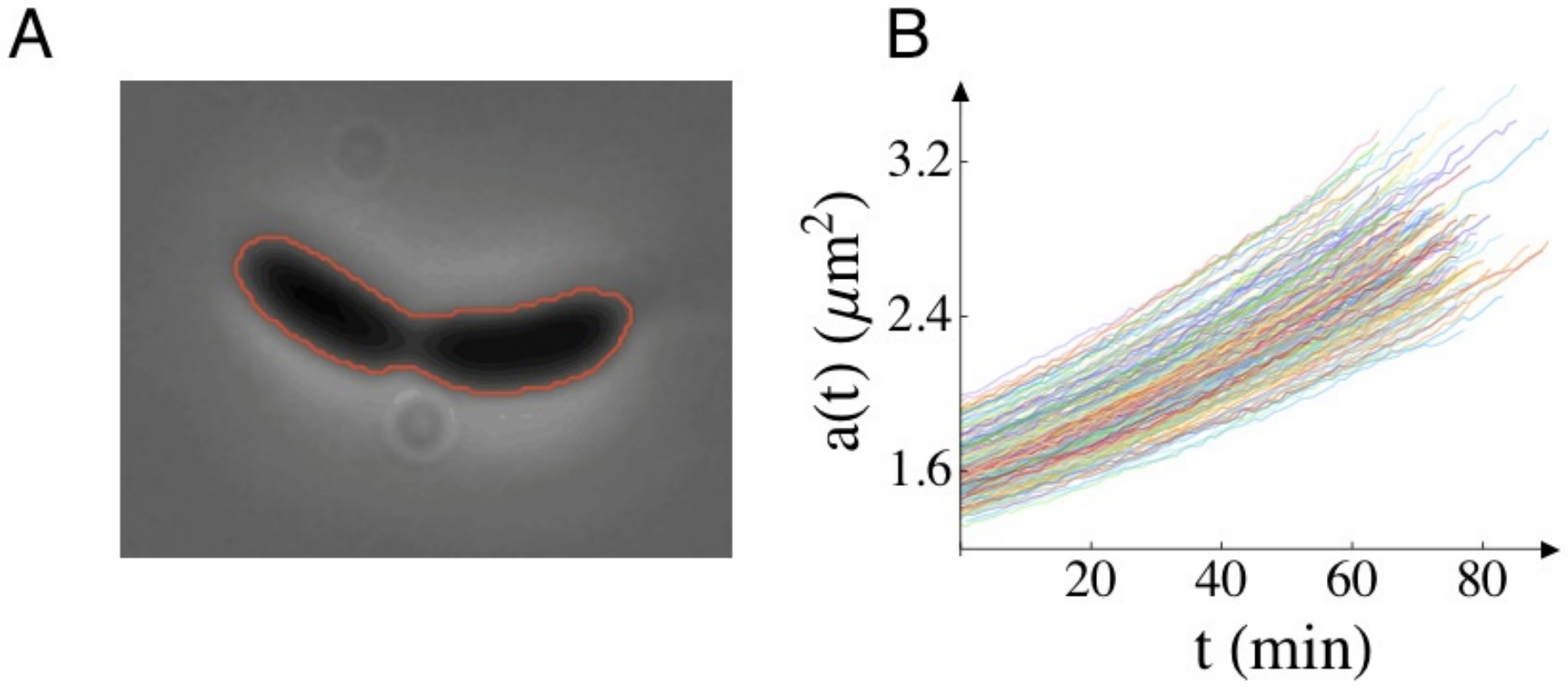}}%inclgphcs[trim=lcm bcm rcm tcm, clip=true, angle=-90]
\caption[From raw images to growth curves]{{\bf From raw images to growth curves.} Phase contrast images such as the one shown in (A) are obtained for each field of view, for each time point, for an experiment at a given temperature. The pixellated boundary (shown in red) of each cell in each frame is extracted by custom image processing algorithms, which combine the absolute intensity level, the spatial gradients of the intensity levels, and a final thresholding step. By linking a sequence of processed images, we obtain area values as a function of time (B), for each generation of each cell. The curves in (B) are plotted with $t$ set equal to $0$ at the beginning of each generation. Data shown are from 5 cells (248 generations total) from an experiment performed at 31$^{\circ}$C.}
%\label{fig-}
\end{center}
\end{figure}
%%%%%%%%%%%%%%%%%%%%%%%end%%%%%%%%%%%%%%%%%%%%%%
%%%%%%%%%%%%%%%%%%%%%%%%begin%%%%%%%%%%%%%%%%%%%%%%%%
\clearpage

\begin{figure}[ht]
\begin{center}
\resizebox{10cm}{!}{\includegraphics[trim=0cm 0cm 0cm 0cm, clip=true]{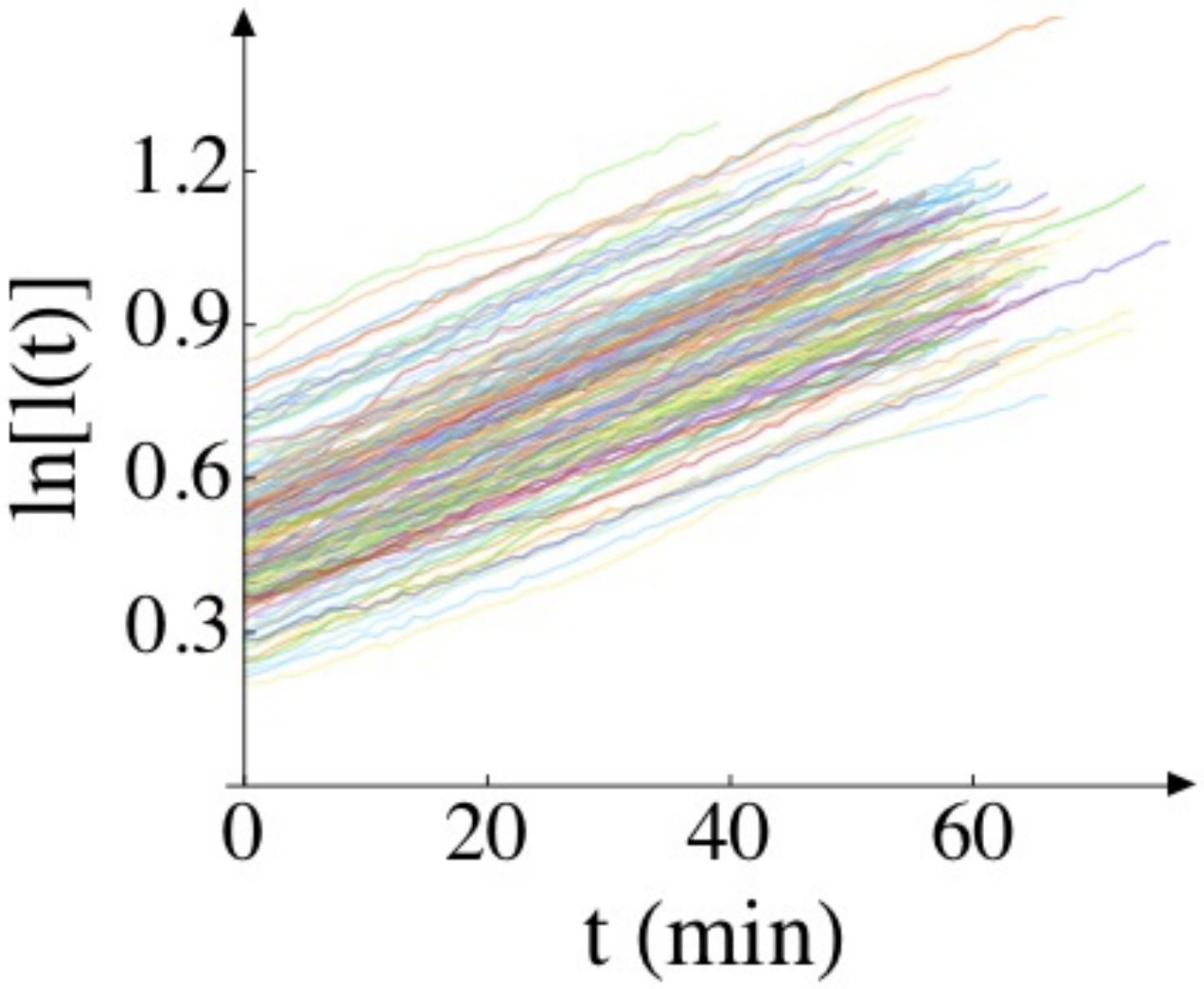}}%[trim=tcm lcm bcm rcm, clip=true, angle=-90]
\caption[Exponential growth of the longitudinal length of the cell]{{\bf Exponential growth of the longitudinal length of the cell.} Data shown are from 5 cells (248 generations total) from an experiment performed at 31$^{\circ}$C. Here we see that $l(t)$, the longitudinal length of the cell,  grows exponentially with time, $t$, as evidenced by the straight lines on the semilog plot shown. ``ln'' stands for the natural logarithm.}
\label{fig-len}
\end{center}
\end{figure}
%%%%%%%%%%%%%%%%%%%%%%%%%end%%%%%%%%%%%%%%%%%%%%%%%%
\clearpage
%%%%%%%%%%%%%%%%%%%%%%begin%%%%%%%%%%%%%%%%%%%%%%
\begin{figure}[ht]
\begin{center}
\resizebox{15cm}{!}{\includegraphics[trim=0cm 6cm 0cm 7cm, clip=true]{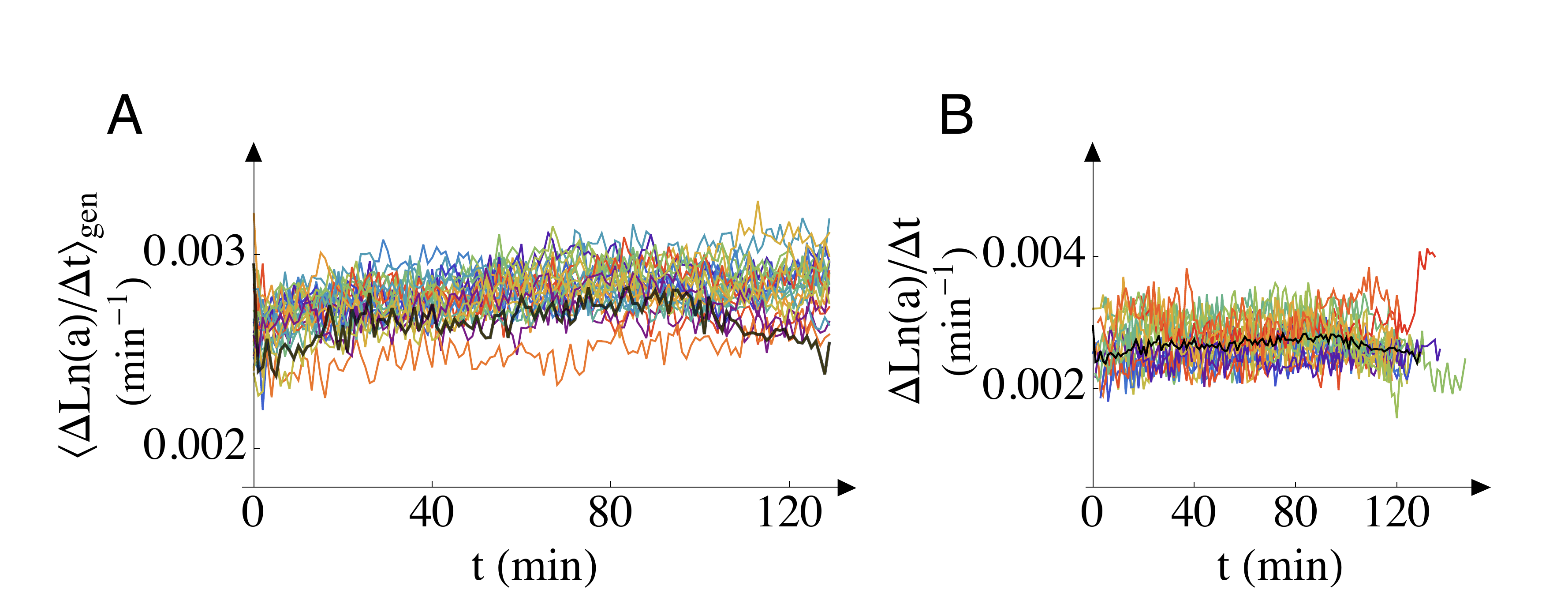}}%inclgphcs[trim=lcm bcm rcm tcm, clip=true, angle=-90]
\caption[Alternative representation of exponential growth]{{\bf Alternative representation of exponential growth.} The rate of change of the logarithmic size is plotted as a function of time. For an exponential growth law, these curves should be parallel to the time axis, and the value of the vertical-axis intercept measures the growth rate, $\k$ for a cell. If the growth law were linear, the slope of this line should change by a factor of 2, which it does not. (A) Each color represents data from one stalked cell, averaged over all its generations. This averaging is denoted by $\la \ldots\ra_{\tiny{gen}}$. Data shown are from the experiment performed at 17$^{\circ}$C.  Since the autocorrelation timescale in the growth curves was estimated to be $\approx$15 min at this temperature, we  consider time points separated by 20 min ($>$correlation time) to evaluate the change in the logarithmic size, so as to ensure statistical independence of successive points. (B) Averaging for a representative cell:  we show the 20 generations that contributed to the black curve in (A). ``Ln'' denotes the natural logarithm.}
\label{fig-s7}
\end{center}
\end{figure}
%%%%%%%%%%%%%%%%%%%%%%%end%%%%%%%%%%%%%%%%%%%%%%
\clearpage

%%%%%%%%%%%%%%%%%%%%%%%%begin%%%%%%%%%%%%%%%%%%%%%%%%
%%%%%%%%%%%%%%%%%%%%%%%%begin%%%%%%%%%%%%%%%%%%%%%%%%
\clearpage
%%%%%%%%%%%%%%%%%%%%%%%%begin%%%%%%%%%%%%%%%%%%%%%%%%
\begin{figure}[ht]
\begin{center}
\resizebox{16cm}{!}{\includegraphics[trim=0cm 8cm 0cm 6cm, clip=true]{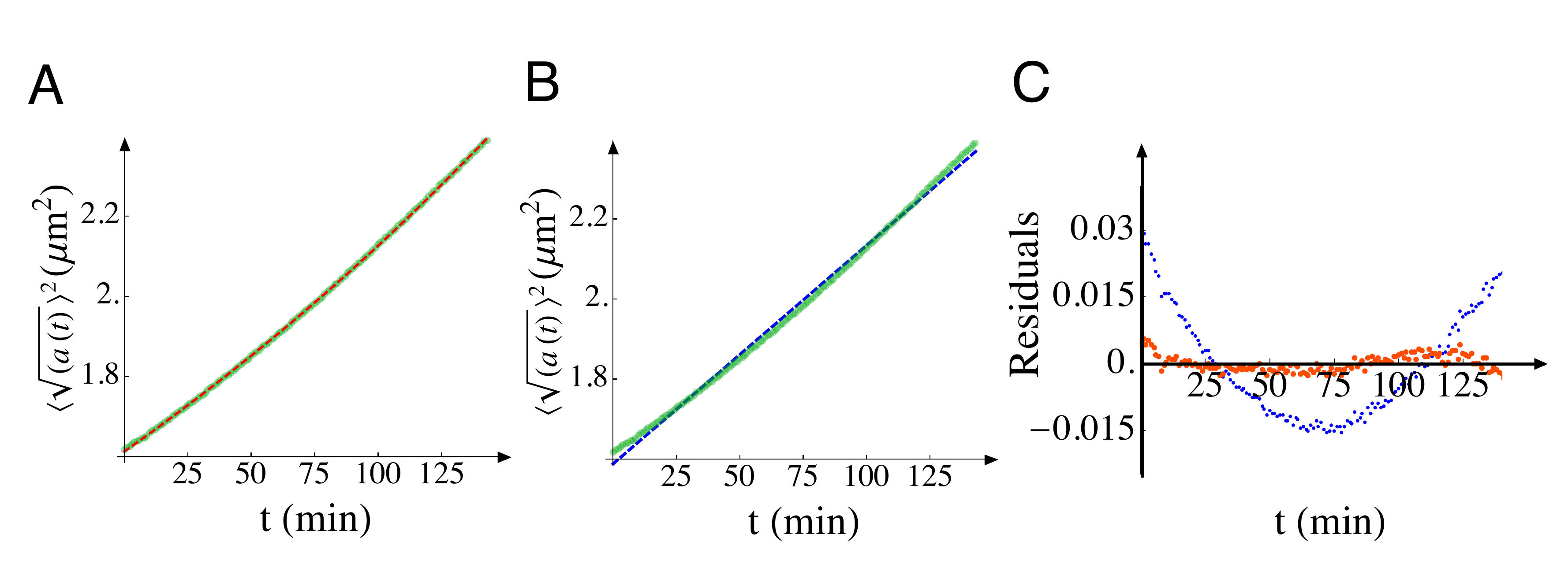}}%[trim=rcm bcm lcm tcm, clip=true, angle=-90
\caption[Exponential vs.\ linear fits for the growth law.]{{\bf Exponential vs.\ linear fits for the growth law.} Experimental data (green) are fit by (A, red) exponential and (B, blue) linear functional forms. (C) Residuals for exponential (red) and linear (blue) fits of the root-mean-square growth curve  for fits in (A) and (B).  Data are for 17$^\circ$C ($\approx$10000 individual growth curves contributing). See SI Text Section \ref{discrim} for discussion. }
\label{fig-exp-growth-S}
\end{center}
\end{figure}
%%%%%%%%%%%%%%%%%%%%%%%%%end%%%%%%%%%%%%%%%%%%%%%%%%
\clearpage
%%%%%%%%%%%%%%%%%%%%%%%%begin%%%%%%%%%%%%%%%%%%%%%%%%
%%%%%%%%%%%%%%%%%%%%%%%%%end%%%%%%%%%%%%%%%%%%%%%%%%
\clearpage

%%%%%%%%%%%%%%%%%%%%%%begin%%%%%%%%%%%%%%%%%%%%%%
\begin{figure}[ht]
\begin{center}
\resizebox{16cm}{!}{\includegraphics[trim=2cm 7cm 1cm 9cm, clip=true]{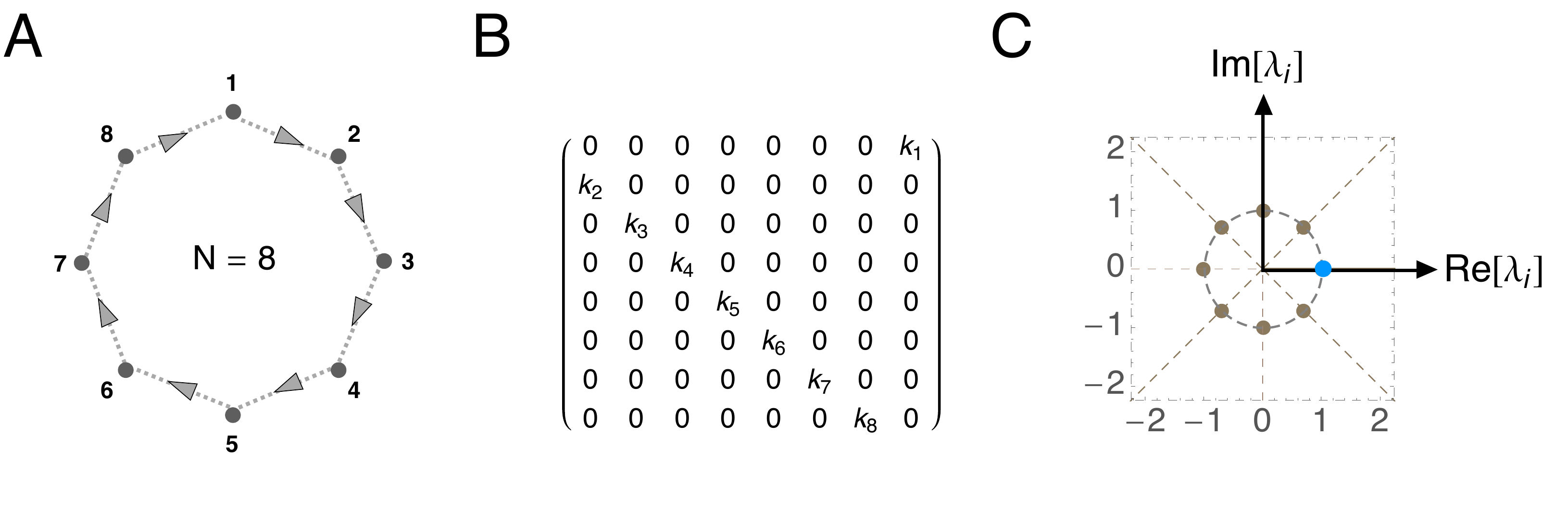}}%inclgphcs[trim=lcm bcm rcm tcm, clip=true, angle=-90]
\caption[]{ {\bf The Hinshelwood cycle yields exponential growth with a rate equal to the geometric mean of constituent rates.} Illustrative example with an $N=8$ Hinshewood cycle (see Fig. 4, main text). (A) Schematic reaction network corresponding to the cycle. (B) The rates can be collected in a matrix, $\mb{K}$.  In this notation \cite{iyer-biswas-fk-s}, each reaction  $X_{i-1}\rightarrow X_{i-1} + X_{i}$ proceeds with rate  $\sum_{j=1}^N\mb{K}_{ij}\, x_{j}$, where $x_i$ is the copy number of species $X_i$, $\mb{K}_{ij} = k_{i}\,\delta_{i-1, j}$, and $\delta$ is the Kronecker delta; the index 0 is equivalent to $N$, closing the cycle.
 (C) The eigenvalues of $\mb{K}$  define the vertices of a regular polygon (here, an octagon, indicated by brown and blue filled circles) in the complex plane. The eigenvalues, $\lambda_i$, are obtained from the roots of the characteristic equation, $\mbox{det}|\mb{K} - \lambda \,\mb{1}| = 0$, or equivalently $\lambda^{8}= k_{1} k_{2} \ldots k_{8}$. Thus there is always only one real positive root (blue), which has a  magnitude equal to $\k$~\cite{iyer-biswas-fk-s}. This eigenvalue dominates the asymptotic dynamics and leads to exponential growth of all $x_{i}$ with growth rate $\k$ (Eq. 2, main text).}
\label{fig-HC}
\end{center}
\end{figure}
%%%%%%%%%%%%%%%%%%%%%%%end%%%%%%%%%%%%%%%%%%%%%%
\clearpage

%%%%%%%%%%%%%%%%%%%%%%begin%%%%%%%%%%%%%%%%%%%%%%
%%%%%%%%%%%%%%%%%%%%%%%end%%%%%%%%%%%%%%%%%%%%%%
\clearpage

%%%%%%%%%%%%%%%%%%%%%%begin%%%%%%%%%%%%%%%%%%%%%%
%%%%%%%%%%%%%%%%%%%%%%%end%%%%%%%%%%%%%%%%%%%%%%
\clearpage
%%%%%%%%%%%%%%%%%%%%%%begin%%%%%%%%%%%%%%%%%%%%%%
%%%%%%%%%%%%%%%%%%%%%%%end%%%%%%%%%%%%%%%%%%%%%%

\end{document}